\documentclass[10pt, sigconf]{acmart}

\usepackage{graphicx}

\setcopyright{none}
\renewcommand\footnotetextcopyrightpermission[1]{} 

\usepackage{booktabs}
\usepackage{algorithmic}
\usepackage{amssymb}
\usepackage{amsfonts}
\usepackage{graphicx}
\usepackage{multicol}
\usepackage{multirow}
\usepackage{psfrag}
\usepackage{subfigure}
\usepackage{verbatim}
\usepackage{epsfig}

\usepackage{algorithm}
\usepackage{algorithm}

\begin{document}

\title[]{Nearly-Linear Time Spectral Graph Reduction for Scalable Graph Partitioning and Data Visualization}

\author{Zhiqiang Zhao}
\affiliation{
\institution{Michigan Technological University}
\city{Houghton}
\state{Michigan}
\postcode{49931}
}
\email{qzzhao@mtu.edu}

\author{Yongyu Wang}
\affiliation{
\institution{Michigan Technological University}
\city{Houghton}
\state{Michigan}
\postcode{49931}
}
\email{yongyuw@mtu.edu }

\author{Zhuo Feng}
\affiliation{
\institution{Michigan Technological University}
\city{Houghton}
\state{Michigan}
\postcode{49931}
}
\email{zhuofeng@mtu.edu}

\begin{abstract}
This paper proposes a scalable algorithmic framework for spectral reduction of large undirected graphs. The proposed method allows computing much smaller graphs while preserving the key spectral (structural) properties of the original graph. Our framework is built upon the following two key components: a spectrum-preserving node aggregation (reduction) scheme, as well as a spectral graph sparsification framework with iterative edge weight scaling. 
We show that the resulting spectrally-reduced graphs  can robustly preserve the first few nontrivial eigenvalues and eigenvectors of the original graph Laplacian.  In addition, the spectral graph reduction method has  been leveraged to develop much faster algorithms for multilevel spectral graph partitioning as well as t-distributed Stochastic Neighbor Embedding (t-SNE) of large data sets.  
We conducted extensive experiments using a variety of large graphs and data sets, and obtained very promising results. For instance, we are able to reduce the ``coPapersCiteseer" graph with $0.43$ million nodes and $16$ million edges to a much smaller graph with only $13K$ (32X fewer) nodes and $17K$ (950X fewer) edges in about 16 seconds; the spectrally-reduced graphs also allow us to achieve up to  $1100X$ speedup for spectral graph partitioning and up to $60X$ speedup for t-SNE visualization of large data sets. 
\end{abstract}

\maketitle

\section{Introduction}\label{sect:introduction}
Recent research shows that by leveraging the key spectral properties of eigenvalues and eigenvectors of graph Laplacians, more efficient algorithms can be developed for tackling many graph-related computing tasks~\cite{teng2016scalable}. For example, spectral methods can potentially lead to much faster algorithms for solving sparse matrices~\cite{spielman2014sdd,zhiqiang:iccad17}, numerical optimization~\cite{christiano2011flow}, data mining~\cite{peng2015partitioning}, graph analytics~\cite{koren2003spectral},  machine learning~\cite{defferrard2016convolutional}, as well as very-large-scale integration (VLSI) computer-aided design (CAD)~\cite{zhuo:dac16,zhiqiang:dac17,zhiqiang:iccad17, zhuo:dac18}.  To this end, {spectral sparsification of graphs} has been extensively studied in the past decade \cite{spielman2011graph,   batson2012twice, spielman2011spectral,Lee:2017}  to allow computing almost-linear-sized \footnote{The number of vertices (nodes) is similar to the number of edges.} subgraphs or {sparsifiers} that can robustly preserve the spectrum, such as the eigenvalues and eigenvectors of the original Laplacian. The sparsified graphs retain the same set of vertices but much fewer edges, which can be regarded as ultra-sparse graph proxies and have been leveraged for developing a series of {nearly-linear-time} numerical and graph algorithms ~\cite{spielman2011graph, Fung:2011stoc, christiano2011flow, spielman2014sdd}.

In this paper, we introduce a scalable algorithmic framework for \emph{spectral reduction of graphs} for dramatically reducing the size (both nodes and edges) of undirected graphs  while preserving the key spectral (structural) properties of the original graph.  
The  spectrally-reduced graphs will immediately lead to the development of much faster numerical and graph-related algorithms. For example, spectrally-reduced social (data) networks may allow for more efficiently modeling, mining and analysis of large social (data) networks, spectrally-reduced neural networks allow for  more scalable model training and processing in emerging machine learning tasks, spectrally-reduced circuit networks may lead to more efficient simulation, optimization and verification of large integrated circuit (IC) systems, etc. 

Our approach consists of two key phases: 1) a  scalable spectrum-preserving node aggregation (reduction) phase, and 2) a spectral graph sparsification phase with iterative subgraph scaling. To achieve truly scalable (nearly-linear time) performance  for spectral graph reduction, we leverage recent similarity-aware spectral graph sparsification method \cite{zhuo:dac18}, graph-theoretic algebraic multigrid (AMG) Laplacian solver \cite{livne2012lean,zhiqiang:iccad17} and a novel constrained stochastic gradient descent (SGD) optimization approach.  The major contribution of this work has been summarized as follows:\\
 \noindent (1) To well preserve the key spectral properties of the original graph in the reduced graph, a nearly-linear time spectrum-preserving node aggregation (reduction) scheme is proposed for robustly constructing reduced graphs that have much less number of nodes.  \\
 
  \noindent (2) A scalable framework for spectral graph sparsification and iterative {subgraph scaling} is introduced for assuring  sparsity  of the reduced graphs by leveraging a novel constrained SGD optimization approach.\\
  
 \noindent (3) We introduce a simple yet effective procedure for refining solutions, such as the Laplacian eigenvectors, computed with spectrally-reduced graphs, which immediately allows using much smaller graphs in many numerical and graph algorithms while achieving superior solution quality. \\
 
  \noindent (4)  In addition, multilevel frameworks that allow us to leverage spectrally-reduced graphs for much faster spectral graph partitioning as well as t-distributed Stochastic Neighbor Embedding (t-SNE) of large data sets are proposed.  \\
  
 \noindent (5) We have obtained very promising experiment results  for a variety of graph problems: the spectrally-reduced graphs  allow  us to achieve up to  $1100X$ speedup for spectral graph partitioning and up to $60X$ speedup for t-SNE visualization of large data sets. \\

The rest of this paper is organized as follows: Section \ref{sect:preliminaries} provides a brief introduction to  spectral graph sparsification; Section \ref{sect:graphreduc} presents the proposed spectral graph reduction approach and its complexity analysis. Section \ref{sect:application} introduces applications of spectral graph reduction methods for scalable graph partitioning and data visualization. Extensive experimental results have been demonstrated in Section \ref{sect:results}, which is followed by the conclusion of this work in Section \ref{sect:conclusion}.

\section{Preliminaries}\label{sect:preliminaries} 
\subsection{Laplacian Matrices of graphs}
 Consider an undirected graph $G=(V,E_G,w_G)$ with $V$ denoting the set of vertices, $E_G$ denoting the set of undirected edges, and $w_G$ denoting the associated edge weights. We define $\mathbf{D_G}$ to be a diagonal matrix with ${D_G}(i,i)$ being equal to the (weighted)  degree of node $i$, and $\mathbf{A_G}$ and $\mathbf{L_G}$ to be the adjacency and Laplacian matrices of undirected graph $G$  as follows, respectively:
\begin{equation}\label{di_laplacian}
\mathbf{A_G}(i,j)=\begin{cases}
w_G(i,j) & \text{ if } (i,j)\in E_G \\
0 & \text{otherwise }
\end{cases}
\end{equation}
Graph Laplacians can be constructed by using $\mathbf{L_G=\mathbf{D_G}-\mathbf{A_G}}$ and will satisfy the following conditions: \textbf{1.} Each column and row sum will be equal to zero; \textbf{2.} All off-diagonal elements are non-positive; \textbf{3.} The graph Laplacian is a symmetric diagonally dominant (SDD) matrix. 

\subsection{Spectral Sparsification of Graphs}
To further push the limit of spectral methods for handling big (data) graphs,  many  research problems for {dramatically simplifying large graphs leveraging spectral graph theory}  have been  extensively studied by mathematics and theoretical computer science (TCS) researchers in the past decade \cite{spielman2011graph, Kolla:2010stoc, batson2012twice, spielman2011spectral, kolev2015note,peng2015partitioning, cohen2017almost,Lee:2017}.  Recent \emph{{spectral graph sparsification }}research allows   constructing nearly-linear-sized  subgraphs that can well preserve the spectral (structural) properties of the original graph, such as the first few eigenvalues and eigenvectors of the graph Laplacian. The related results can potentially lead to the development of a variety of {\emph{nearly-linear time}}  numerical and graph algorithms for solving large sparse  matrices, graph-based semi-supervised learning (SSL), computing the  stationary distributions of Markov chains and personalized PageRank vectors, spectral graph partitioning and data clustering, max-flow of undirected graphs, etc \cite{Spielman:usg,spielman2011graph,spielman2010algorithms,Kolla:2010stoc,miller:2010focs, Fung:2011stoc,spielman2011spectral, christiano2011flow, spielman2014sdd,cohen2017almost}. 

Spectral graph sparsification aims to find a spectrally-similar subgraph (sparsifier) $P=(V,E_P,w_P)$ that has the same set of vertices of the original graph $G=(V,E_G,w_G)$, but  much fewer edges.  There are two types of  sparsification methods: the  cut sparsification methods  preserve the  cuts of the original graph through random sampling of edges \cite{benczur1996approximating}, whereas spectral sparsification methods preserve the graph spectral (structural) properties, such as distances between vertices,  cuts in the graph, as well as the  stationary distributions of Markov chains \cite{cohen2017almost,spielman2011spectral}.  Therefore, spectral graph sparsification is a much stronger notion than cut sparsification. We say  $G$ and its subgraph $P$ are {$\sigma-$spectrally similar} if the following condition holds for all real vectors $\mathbf{x} \in \mathbb{R}^V$: 
\begin{equation}
\label{formula_spectral_similar}
\frac{\mathbf{x}^\top{\mathbf{L}_{P}}\mathbf{x}}{\sigma}\le \mathbf{x}^\top{\mathbf{L}_G}\mathbf{x} \le \sigma \mathbf{x}^\top{\mathbf{L}_{P}}\mathbf{x},
\end{equation}
where $\mathbf{L}_{G}$ and $\mathbf{L}_{P}$ denote the Laplacian matrices  of graph $G$ and $P$, respectively. Define the relative condition number as $\kappa({\mathbf{L}_G},{\mathbf{L}_{P}})=\lambda_{\max}/\lambda_{\min}$, where $\lambda_{\max}$ and $\lambda_{\min}$ are the largest and smallest nonzero eigenvalues of
\begin{equation}
\label{formula_eig_perturb0}
\mathbf{L}_G\mathbf{u}=\lambda \mathbf{L}_{P}\mathbf{u},
\end{equation}
where $\mathbf{u}$ is the generalized eigenvector of $\mathbf{L_G}$. It can be further shown that $\kappa({\mathbf{L}_G},{\mathbf{L}_{P}})\le\sigma^2$, which indicates that a smaller relative condition number or $\sigma^2$ corresponds to a higher spectral similarity. Recent nearly-linear time spectral sparsification algorithm leverages spectral perturbation analysis  to construct nearly-linear-sized spectrally-similar subgraphs \cite{zhuo:dac16,zhuo:dac18}, which leads to the development of much  faster SDD matrix solvers \cite{zhiqiang:iccad17} and spectral graph partitioning algorithm \cite{zhuo:dac18}.


\section{Spectral Reduction of Graphs}\label{sect:graphreduc}
\subsection{Overview}
\begin{figure}[htb]
\begin{center}
	\includegraphics[scale=0.39]{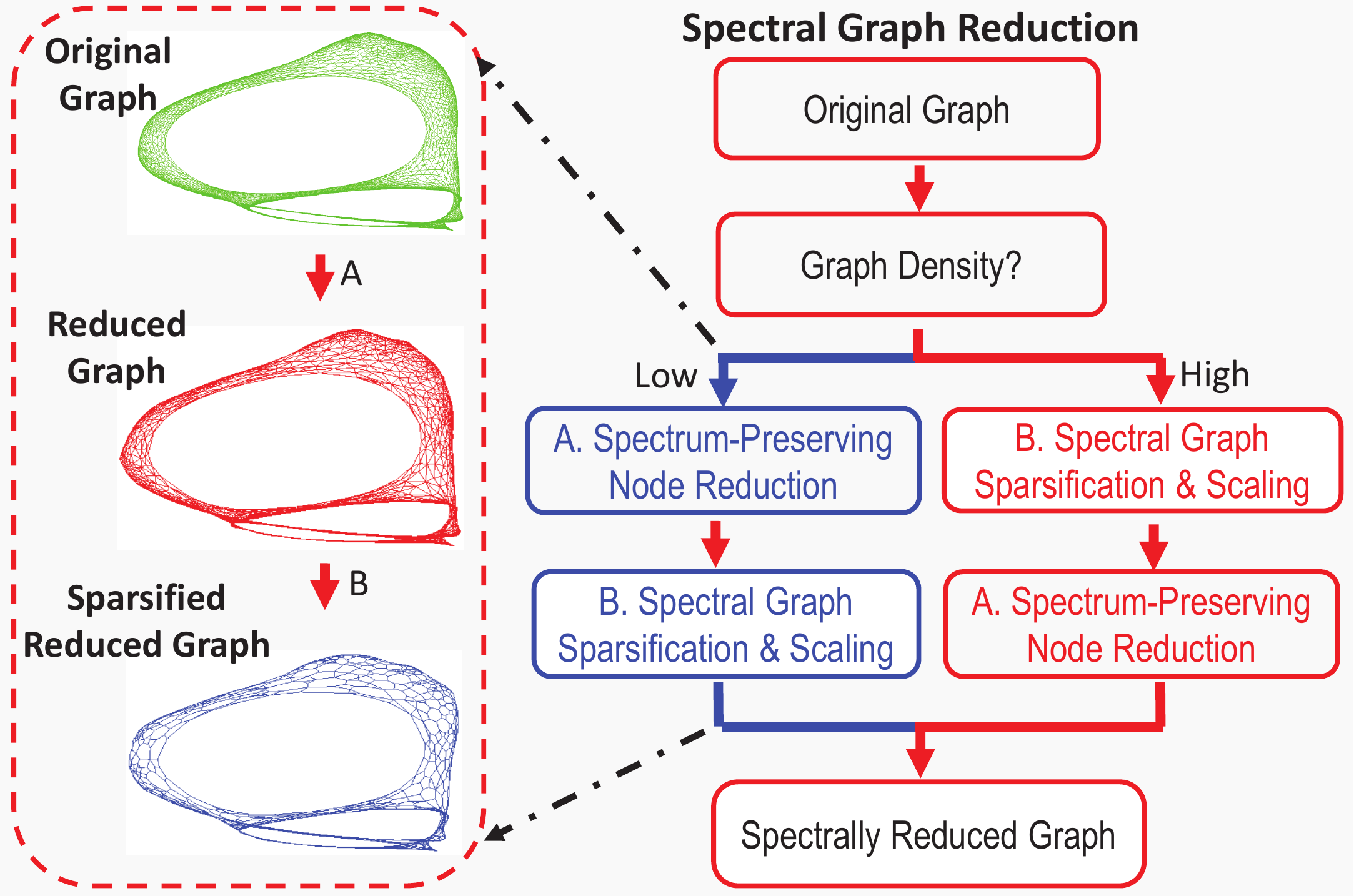}
	\caption{The proposed spectral  reduction framework. \protect\label{fig:flow}}
\end{center}
\end{figure}

In the following,  assume that $G=(V,E_G,w_{G})$ is a weighted, undirected, and connected graph, $P = (V, E_{P}, w_{P})$ is the spectrally sparsified graph of $G$, $R = (V_{R}, E_{R}, w_{R})$ is the reduced graph of $G$ without sparsification, and $S = (V_{R}, E_{S}, w_{S})$ is the  sparsified reduced graph of $G$. The Laplacian matrices of the corresponding graphs have been shown in Table \ref{table:symbol} that also includes the fine-to-coarse (G-to-R) graph mapping matrix denoted by $\mathbf{H_G^R}$ as well as the coarse-to-fine (R-to-G) graph mapping matrix denoted by $\mathbf{H_R^G}$.
 
This work introduces a {\textit{spectral graph reduction}} framework (as shown in Figure \ref{fig:flow}) that allows computing much smaller yet spectrally-similar graph $S$ such that the following condition holds for all real vectors $\mathbf{x} \in \mathbb{R^V}$: 
\begin{equation}
\label{formula_spectral_similar}
\frac{\mathbf{x_R}^\top{\mathbf{L}_{S}}\mathbf{x_R}}{\sigma}\le \mathbf{x}^\top{\mathbf{L}_G}\mathbf{x} \le \sigma \mathbf{x_R}^\top{\mathbf{L}_{S}}\mathbf{x_R},~~~\mathbf{x_R=H_G^R x}.
\end{equation}
 
\begin{table}
\scriptsize\centering \caption{Symbols and their denotations in this work}
\begin{tabular}{ |c|c|c|c|  }
 \hline
 \hline
 Symbol& Denotation &Symbol& Denotation\\
 \hline
 $G = (V, E_{G}, w_{G})$   & The Original  Graph    &$\mathbf{L_{G}}$& Lap. of  $G$\\
 $P = (V, E_{P}, w_{P})$   & Spectrally-Spar. $G$  & $\mathbf{L_{P}}$  &Lap. of  $P$\\
 $R = (V_{R}, E_{R}, w_{R})$ & Reduced $G$ w/o spar.&  $\mathbf{L_{R}}$ &Lap.  of  $R$ \\
 $S = (V_{R}, E_{S}, w_{S})$&  Reduced $G$ w/ spar.& $\mathbf{L_{S}}$&Lap.   of  $S$\\
  $\mathbf{H_G^R}\in  \mathbb{R}^{V_R\times V}$ & G-to-R mapping & $\mathbf{H_R^G}\in  \mathbb{R}^{V\times V_R}$& R-to-G mapping\\
 \hline
 \end{tabular}\label{table:symbol}
\end{table}

An overview of the proposed method for spectral reduction of large graphs is described as follows. Our approach for spectral reduction of undirected graphs includes the following two phases:  \textbf{Phase (A)} will determine  the fine-to-coarse graph mapping operator using spectral node proximity measurement computed based on algebraic distance  \cite{chen2011algebraic}, and reduce the original graph into a much smaller graph using the fine-to-coarse graph mapping operator; \textbf{Phase (B)} will extract spectrally-similar sparsifiers of the original (reduced) graph and scale up edge weights in the sparsified graphs to better match the key spectral (structural) properties, such as the eigenvalues/eigenvectors of graph Laplacians. Since the spectral node proximity metric based on algebraic distance cannot be directly applied to dense graphs \cite{chen2011algebraic}, our approach will first examine the average  node degrees in the original graph: if the original graph is relatively sparse ($|E_G|<40|V|$), \textbf{Phases (A) to (B)} will be performed in sequence as shown in Figure \ref{fig:flow}; otherwise, if the original graph is too dense ($|E_G|>40|V|$), \textbf{Phase (B)} for  spectral graph sparsification and edge scaling will be performed first, which is followed by \textbf{Phase (A)}.

\subsection{Phase (A): Spectrum-Preserving Node Reduction }\label{sec:node_redu}
\subsubsection{Spectral node affinity metric.}
To generate the reduced graph based on the original graph,  a spectrum-preserving node aggregation scheme is applied based on spectral node affinity    $a_{p,q}$ defined as follows for neighboring nodes $p$ and $q$ \cite{livne2012lean,chen2011algebraic}:
\begin{equation}\label{eqn:agg}
a_{p,q} = \mathbf{\frac{{\|(X_p, X_q)\|}^2}{(X_p, X_p)(X_q, X_q)},   (X_p,X_q) = \Sigma_{k=1}^{K}{(x_p^{(k)}\cdot x_q^{(k)})}}
\end{equation}
where $\mathbf{X = (x^{(1)},\dots, x{^{(K)}})}$ includes $K$ test vectors computed by applying a few Gauss-Seidel (GS) relaxations for solving the linear system of equations $\mathbf{L_{G} x^{(i)}}=0$ for $i=1,...,K$ with $K$ random vectors that are orthogonal to the all-one vector $\mathbf{1}$ or equivalently satisfying $\mathbf{1^\top x^{(i)}=0}$. Let $\mathbf{\tilde{x}^{(i)}}$ denote the approximation of the true solution $\mathbf{x^{(i)}}$ after applying several GS relaxations to $\mathbf{L_{G} x^{(i)}}=0$. Due to the smoothing property of GS relaxation, the latest error can be expressed as $\mathbf{e^{(i)}_s \,=\,x^{(i)}-\tilde{x}^{(i)}}$, which will only contain the smooth (low-frequency) modes of the initial error, while the oscillatory (high-frequency) modes of the initial error will be effectively eliminated \cite{briggs2000multigrid}.  Based on the $K$ smoothed vectors in $\mathbf{X}$, it is possible to embed each node into a $K$-dimensional space such that node $p$ and node $q$ are considered spectrally-close enough to each other if their low-dimensional embedding vectors $\mathbf{x_p}\in \mathbb{R}^K$ and $\mathbf{x_q}\in \mathbb{R}^K$ are highly correlated. Spectrally-similar nodes $p$ and  $q$ can be then aggregated together for node reduction purpose.

\subsubsection{Spectral similarity between nodes.}
It has been shown that the node affinity metric $a_{p,q}$ can usually effectively reflect the distance or strength of connection between nodes $p$ and  $q$ in a graph \cite{livne2012lean}: a larger $a_{p,q}$ value indicates a stronger spectral similarity (correlation) between nodes $p$ and $q$.  The above affinity metric can be better understood by looking at the algebraic distance $d_{p,q}$ between node $p$ and $q$ computed by $d_{p,q}=1-a_{p,q}$ that can be used to represent the geometric distance in grid-structured graphs. For example, consider an $n$-node $1-D$ path graph with a discretization size of $h=1/n$. It can be shown that $d_{p,q}$ is proportional to $(p-q)^2*h^2$ in the discretized Poisson equation in which nodes locate at $t*h$, for $t=0,1\dots,n$ and $h\Rightarrow 0$. Consequently, the nodes with large affinity or small algebraic distance should be aggregated together to form the nodes in the reduced graph. Once node aggregation schemes are determined, the  graph mapping operators ($\mathbf{H_G^R}$ and $\mathbf{H_R^G}$) can be obtained and leveraged for constructing spectrally-reduced graphs. For example, the reduced Laplacian can be computed by $\mathbf{L_R=H_G^R L_G H_R^G}$, which uniquely defines the reduced graph. 

We emphasize that the node aggregation  (reduction) scheme based on the above spectral node affinity calculations will have a (linear) complexity of $O(|E_G|)$ and thus allow preserving the spectral (global or structural) properties  of the original graph in the reduced graph in a highly efficient and effective way: the smooth components in the first few Laplacian eigenvectors can be well preserved after node aggregation, which is key to preserving the first few (bottom) eigenvalues and eigenvectors of the original graph Laplacian in the reduced graphs \cite{pmlr-v80-loukas18a}. 

\subsubsection{Limitations when dealing with dense graphs.}
The above node reduction scheme based on the algebraic distance metric may not be reliable when applied to dense graph problems. Since each node in the dense graph will typically connect to many other nodes, running a few GS relaxations will result in many nodes seemingly close to each other and can lead to rather poor node aggregation results. For example, an extreme case is to directly apply the above node aggregation scheme to a complete graph where each node  has $|V|-1$ edges connecting to the rest of the nodes: since applying  GS relaxations will immediately assign same values to all nodes, no meaningful clusters of nodes can be identified. As shown in our experiment results, it is not possible to use the above node affinity metric for aggregating nodes for the ``appu" graph \cite{davis2011matrix} that has high average node degrees ($|E_G|/|V|\approx 90$). 

To this end, we propose to perform a spectral sparsification and scaling procedure (\textbf{Phase (B)}) before applying the node aggregation (reduction) phase. Such a scheme will allow extracting  ultra-sparse yet spectrally-similar subgraphs and subsequently aggregate nodes into clusters using  the above node affinity metric. As a result, the spectral graph reduction flow proposed in this work can be reliably applied to handle both sparse and dense graphs, as shown in Figure \ref{fig:flow}.

\subsection{Phase (B): Spectral Graph Sparsification and Scaling}
\label{sec:top_spar}
\subsubsection{Copping with high graph densities.}
The proposed spectral node aggregation scheme in Section \ref{sec:node_redu} will enable us to reliably construct smaller graphs that have much less number of nodes. However, the aggregated nodes may potentially result in much denser graphs (with significantly higher node degrees), which may  incur even greater computational and memory cost for graph operations. For example, emerging multi-way spectral graph partitioning (clustering) algorithms \cite{lee2014multiway,peng2015partitioning} require to compute multiple Laplacian eigenvectors, which can   still be very costly for   dense graphs  since the efficiency of nowadays eigen-solvers or eigendecomposition methods strongly depend on the matrix sparsity \cite{xue2009numerical,lehoucq1997arpack,saad2003iterative}.

To address the challenging issues caused by relatively dense graphs,  we propose the following highly effective yet scalable algorithms in \textbf{Phase (B)}: the nearly-linear time spectral graph sparsification and subgraph  scaling schemes for handling dense graphs $G$. Note that when \textbf{Phase (B)} is applied for a sparse input graph, the same procedures can be applied to the reduced graph $R$ (with potentially higher density) for computing $S$ after the node aggregation scheme or the fine-to-coarse graph mapping operator is determined.

\subsubsection{Spectral approximation with spanning tree subgraphs.}
Denote the total stretch of the spanning-tree subgraph $P$ with respect to the original graph $G$ to be $\textstyle{{\rm{st}}}_{P}(G)$. Spielman~\cite{spielman2009note} showed that $\mathbf{L}^+_{P} \mathbf{L}_G$ has at most $k$ generalized eigenvalues greater than $\textstyle{{\rm{st}}}_{P}(G)/k$. It has been shown that every graph has a { low-stretch spanning tree} (LSST) with bounded total stretch \cite{spielman2009note}, which leads to:
 \begin{equation}\label{formula_stretch}
\kappa({\mathbf{L}_G},{\mathbf{L}_{P}}) \leq \textstyle{{\rm{Tr}}}({\mathbf{L}^+_{P} \mathbf{L}_G}) = \textstyle{{\rm{st}}}_{P}(G) \leq (m \log n \log \log n),
\end{equation}
where $m=|E_G|$, $n=|V|$, and $\textstyle{{\rm{Tr}}}({\mathbf{L}^+_{P} \mathbf{L}_G})$ is the trace of ${\mathbf{L}^+_{P} \mathbf{L}_G}$. Such a result motivates to construct an ultra-sparse yet spectrally-similar subgraphs by recovering only a small portion of important off-tree edges to the spanning tree. For example, a recent spectral perturbation framework \cite{zhuo:dac16,zhuo:dac18} allows  constructing the $\sigma$-similar spectral sparsifiers with $O(m \log n \log \log n/\sigma^2)$ off-tree edges in nearly-linear time.

\subsubsection{Towards better approximation with off-tree edges.}
To dramatically reduce the spectral distortion (the total stretch) between the original graph and the spanning tree,  a spectral off-tree edge embedding scheme and edge filtering method with approximate generalized eigenvectors have been proposed in \cite{zhuo:dac16,zhuo:dac18}, which is based on following spectral perturbation analysis:
\begin{equation}\label{formula_eig_perturb1}
\mathbf{L}_G({\mathbf{u}_i + \delta \mathbf{u}_i}) = ({\lambda _i + \delta \lambda _i})(\mathbf{L}_{P} + \delta \mathbf{L}_{P})({\mathbf{u}_i + \delta \mathbf{u}_i}),
\end{equation}
where a perturbation $\delta \mathbf{L}_{P}$ is applied to $\mathbf{L}_{P}$, which results in perturbations in generalized eigenvalues ${\lambda _i} + \delta {\lambda _i}$ and eigenvectors ${\mathbf{u}_i} + \delta {\mathbf{u}_i}$ for $i=1, \ldots, n$, respectively. The first-order perturbation analysis \cite{zhuo:dac16} leads to:
\begin{equation}\label{formula_eig_perturb5}
-\frac{\delta {\lambda _i}}{{\lambda _i}} = \mathbf{u}_i^\top\delta \mathbf{L}_{P}{\mathbf{u}_i},
\end{equation}
which indicates that the reduction of $\lambda _i$ is proportional to the Laplacian quadratic form of $\delta \mathbf{L}_{P}$ with the generalized eigenvector $\mathbf{u}_i$. Therefore, if the dominant eigenvector $\mathbf{u}_n$ is applied, the largest generalized eigenvalue $\lambda_n$ can be significantly reduced by properly choosing $\delta \mathbf{L}_{P}$ that includes the set of off-tree edges and their weights. Once the largest generalized eigenvalue becomes sufficiently small, the distortion between subgraph $P$ and original graph $G$ will be greatly reduced.  

An alternative view of such a spectral embedding scheme is to consider the following \textbf{Courant-Fischer theorem } for generalized eigenvalue problems:
\begin{equation}\label{formula_courant-fischer-max}
\lambda_{n}=\mathop{\max_{|\mathbf{x}|\neq 0}}_{\mathbf{x}^\top\mathbf{1}=0}\frac{\mathbf{x}^\top \mathbf{L}_G \mathbf{x}}{\mathbf{x}^\top \mathbf{L}_P \mathbf{x}}\approx\mathop{\max_{|\mathbf{x}|\neq 0}}_{x(p)\in \left\{0,1 \right\}}\frac{\mathbf{x}^\top \mathbf{L}_G \mathbf{x}}{\mathbf{x}^\top \mathbf{L}_P \mathbf{x}}=\mathop{\max}\frac{|\partial_G (Q)|}{|\partial_P (Q)|},
\end{equation}
where  $\mathbf{1}$ is the all-one vector, the node set $Q$ is defined as
\begin{equation}
Q \overset{\mathrm{def}}{=}\left\{ p \in V: x(p)=1 \right\},
\end{equation}
 and the boundary of $Q$ in $G$ is defined as
\begin{equation}
\partial_G (Q) \overset{\mathrm{def}}{=}\left\{ (p,q)\in E_G: p\in Q, q\notin Q\right\},
\end{equation} 
which will lead to
\begin{equation}
\begin{split}
&\mathbf{x}^\top \mathbf{L}_G \mathbf{x}=|\partial_G (Q)|,\\
&\mathbf{x}^\top \mathbf{L}_P \mathbf{x}=|\partial_P (Q)|.
\end{split}
\end{equation}
 According to (\ref{formula_courant-fischer-max}), $\lambda_{max}=\lambda_{n}$ will reflect the largest mismatch of the boundary (cut) size between $G$ and $P$, since finding the dominant generalized eigenvector is similar to finding the node set $Q$ such that $\frac{|\partial_G (Q)|}{|\partial_P(Q)|}$ or the mismatch of boundary (cut) size between the original graph $G$ and subgraph $P$ is maximized.  Once $Q$ or $\partial_P(Q)$ can be identified by spectral graph embedding using dominant generalized eigenvectors, the edges in $\partial_G (Q)$ can be selected and recovered to $P$ to dramatically reduce the maximum mismatch or $\lambda_{n}$. 

Denote $\mathbf{e}_{p}\in \mathbb{R}^V$ to be the elementary unit vector with only the $p$-th element being $1$ and others being $0$, and we denote $\mathbf{e}_{p,q}=\mathbf{e}_{p}-\mathbf{e}_{q}$. Then by including the off-tree edges, the generalized eigenvalue perturbation can be expressed as follows:
\begin{equation}\label{formula_eig_perturb6}
-\frac{\delta {\lambda _i}}{{\lambda _i}}=\mathbf{u}_i^\top \delta \mathbf{L}_{P,max}\mathbf{u}_i = \sum_{(p,q)\in E_G\setminus E_P}^{}{{w_{G}(p,q)}\left(\mathbf{e}_{p,q}^T\mathbf{u}_i\right)^2},
\end{equation}
where $\delta\mathbf{L}_{P,max}=\mathbf{L}_{G}-\mathbf{L}_{P}$, and $ w_G({p,q})$ denotes the weight of edge $(p,q)$ in the original graph. The \textbf{spectral criticality} $c_{p,q}$ of each off-tree edge $(p,q)$ is defined as:
 \begin{equation}\label{formula_criticality}
c_{p,q}={{w_{G}(p,q)}\left(\mathbf{e}_{p,q}^T\mathbf{u}_n\right)^2}.
\end{equation}
If we consider the undirected graph $G$ to be a resistor network, and  $\mathbf{u}_n$ to be the voltage vector for that resistor network, $c_{p,q}$ can be regarded as the edge Joule heat (power dissipation). Consequently, the most spectrally critical off-tree edges from $\partial_G(Q)$ can be identified and recovered into LSST for  spectral graph topology sparsification by (\ref{formula_criticality}), which allows improving spectral approximation in the subgraph by dramatically reducing the $\lambda_{n}$. In practice, approximate generalized eigenvectors computed through a small number of generalized power iterations will suffice for low-dimensional spectral off-tree edge embedding, which can be realized as follows:\\

\noindent\textbf{(1)} Compute an approximate eigenvector $\mathbf{h}_t$ by applying $t$-step generalized power iterations on an initial vector $\mathbf{h}_0=\sum\limits_{i = 1}^{n} {{\alpha _i}{}{\mathbf{u}_i}}$:
\begin{equation}\label{formula_pwr_iter5}
{\mathbf{h}_t}= \left({\mathbf{L}_{P}^{+}}\mathbf{L}_G\right)^t{\mathbf{h}_0} =\left(\sum_{i=1}^{n}{\lambda_i\mathbf{u}_i\mathbf{u}_i^T}\right)^t\sum_{i = 1}^{n} {{\alpha _i}{}{\mathbf{u}_i}}= \sum_{i = 1}^{n} {{\alpha _i}{\lambda^t _i}{\mathbf{u}_i}}; 
\end{equation}
\noindent\textbf{(2)} Compute the quadratic form for off-tree edges with $\mathbf{h_t}$:
  \begin{equation}\label{formula_pwr_iter5new}
  \begin{array}{l}
-\frac{\delta {\lambda _i}}{{\lambda _i}}\approx \mathbf{h_t}^\top \delta\mathbf{L}_{P,max} \mathbf{h_t}=\sum\limits_{i = 1}^{n} {{{\left( {{\alpha _i}{\lambda^t _i}} \right)}^2(\lambda_i-1)}}\\=\sum\limits_{({p,q})\in E_G\setminus E_P}^{} w_G(p,q)\left(\mathbf{e}_{p,q}^T\mathbf{h}_t\right)^2=\sum\limits_{({p,q})\in E_G\setminus E_P}^{}\tilde c_{p,q},
\end{array}
\end{equation}
where $\tilde c_{p,q}$ denotes the {approximate spectral criticality} of each off-tree edge $(p,q)$.
It should be noted that using $k$ vectors computed by (\ref{formula_pwr_iter5}) will enable  to embed each node into a $k$-dimensional \textbf{generalized eigenspace}, which can facilitate edge filtering from $\partial_G (Q)$  to avoid recovering similar edges into $P$. In this work, we choose $t=2$, which already leads to consistently good results for a large variety of graph problems. To achieve more effective edge filtering for  similarity-aware spectral graph sparsification, an incremental graph densification procedure \citep{zhuo:dac18} will be adopted in this work. During each graph densification iteration, a small portion of ``filtered" off-tree edges will be added to the latest spectral sparsifier, while the  spectral similarity is estimated to determine if more off-tree edges are needed.

\subsubsection{Subgraph scaling via constrained  optimization.}
To aggressively limit the number of edges in the subgraph $P$ while still achieving a high quality approximation of the original graph $G$, we propose an efficient spectral scaling scheme for scaling up edge weights in the subgraph $P$ to further reduce the largest mismatch or $\lambda_n$. The dominant eigenvalue perturbation $\delta \lambda_n$ can be expressed in terms of edge weight perturbations as follows:
 \begin{equation}\label{formula_weight_sensitivity}
-\frac{\delta {\lambda _n}}{{\lambda _n}}= \mathbf{u}^\top_n\delta \mathbf{L}_{P}\mathbf{u}_n = \sum_{({p,q})\in E_P}^{} {{\delta w_P(p,q)}\left(\mathbf{e}_{p,q}^\top\mathbf{u}_n\right)^2},
\end{equation}
\noindent which directly gives the sensitivity of $\lambda _n$ with respect to each edge weight $w_{P}(p,q)$ in graph $P$:
 \begin{equation}\label{formula_weight_sensitivity2}
  \frac{\delta \lambda _n}{\delta w_P(p,q)}=-\lambda _n \left(\mathbf{e}_{p,q}^\top\mathbf{u}_n\right)^2\approx -\lambda _n \left(\mathbf{e}_{p,q}^\top\mathbf{h}_t\right)^2.
\end{equation}
 \noindent The (approximate) sensitivity expressed in (\ref{formula_weight_sensitivity2}) can be leveraged for finding a proper edge weight scaling factor for each edge in $P$ such that $\lambda_n$ will be dramatically reduced. Since scaling up edge weights in $P$ will result in the monotonic decrease of both $\lambda_n$ and $\lambda_1$, it is likely that $\lambda_1$ will decrease at a faster rate than $\lambda_n$, which leads to a degraded spectral similarity between $G$ and $P$. To avoid such a degradation in spectral approximation quality, we propose the following methods for estimating the extreme generalized eigenvalues $\lambda_n$ and $\lambda_1$, which allows us to more properly scale up edge weights in $P$.

The largest eigenvalues of $\mathbf{L}^+_{P} \mathbf{L}_G$ are well separated from each other \cite{spielman2009note}, which allows computing rather accurate  largest eigenvalue ($\lambda_{n}$) by performing only a small number of generalized power iterations with an initial random vector. Since the generalized power iterations can converge at a geometric rate governed by $\lambda_{n-1}/\lambda_n$,  the error of the estimated largest generalized eigenvalue  will drop to $|\lambda_{n-1}/\lambda_n|^k e_0$ after $k$ iterations for an initial error $e_0$. As a result, only a few (five to ten) iterations will be sufficient to compute a good estimation of $\lambda_n$ for well separated largest eigenvalues that lead to small $\lambda_{n-1}/\lambda_n$.  To gain scalable runtime performance, we  will leverage  recent graph-theoretic algebraic multigrid (AMG) algorithms for  solving the sparsified Laplacian matrix $\mathbf{L_P}$  \cite{livne2012lean, zhiqiang:iccad17}.

Since the smallest eigenvalues of $\mathbf{L}^+_{P} \mathbf{L}_G$ are crowded together, using (shifted) inverse power iterations may not be efficient due to the slow convergence caused by relatively poor separation of smallest eigenvalues. To more efficiently estimate the smallest generalized eigenvalue, we leverage the Courant-Fischer theorem for approximately computing the  smallest generalized eigenvalues: 
\begin{equation}\label{formula_courant-fischer}
\lambda_1=\lambda_{min}=\mathop{\min_{|\mathbf{x}|\neq 0}}_{\mathbf{x}^\top\mathbf{1}=0}\frac{\mathbf{x}^\top \mathbf{L}_G \mathbf{x}}{\mathbf{x}^\top \mathbf{L}_P \mathbf{x}},
\end{equation}
which indicates that the key to locating the smallest generalized eigenvalues is to find a vector $\mathbf{x}$ that minimizes the ratio between the quadratic forms of the original and sparsified Laplacians. In our method, we will require  every element in  $\mathbf{x}$ to only take a value $1$ or $0$ for each node in both $G$ and $P$ for minimizing the following ratio, which will lead  to a good estimation for $\lambda_1$:
\begin{equation}\label{formula_courant-fischer-2}
\lambda_1\approx\mathop{\min_{|\mathbf{x}|\neq 0}}_{x(p)\in \left\{0,1 \right\}}\frac{\mathbf{x}^\top \mathbf{L}_G \mathbf{x}}{\mathbf{x}^\top \mathbf{L}_P \mathbf{x}}=\mathop{\min_{|\mathbf{x}|\neq 0}}_{x(p)\in \left\{0,1 \right\}}{\frac{\sum\limits_{x(p)\neq x(q), (p,q)\in E_G} w_G(p,q)}{\sum\limits_{x(p)\neq x(q), (p,q)\in E_P} {w}_P(p,q)}},
\end{equation}
To this end, we initialize all nodes with the same value of $0$  and only select a single node $p$ to be assigned with a value of $1$, which leads to:
\begin{equation}\label{formula_courant-fischer-3}
\lambda_{1}\approx \min_{p\in V}{\frac{ \mathbf{d}_G(p)}{\mathbf{d}_P(p)}},
\end{equation}
where $\mathbf{d}_G$ and $\mathbf{d}_P$ are the diagonal vectors of $\mathbf{L_G}$ and $\mathbf{L_P}$ satisfying $\mathbf{d}_G(p)=\mathbf{L}_G(p,p)$ and $\mathbf{d}_P(p)=\mathbf{L}_P(p,p)$.  (\ref{formula_courant-fischer-3}) indicates that $\lambda_{1}$ can be well approximated in linear time by finding the node with minimum weighted degree ratio of $G$ and $P$. 
 
 Based on the above scalable methods for estimating the extreme eigenvalues $\lambda_{1}$ and $\lambda_{n}$ of   $\mathbf{L}^+_{P} \mathbf{L}_G$, as well as the weight sensitivity in  (\ref{formula_weight_sensitivity2}),  the following constrained nonlinear optimization framework for scaling up edge weights in  $P$ has been proposed.
\begin{equation}\label{edge_opt}
\begin{split}
  {~~~\textbf{minimize:}} ~~~~~ \lambda_{n}(w_P)\\
 {\textbf{s. t.:}}~~~~~~~~~~~~~~~~\\
\textbf{(a)}~~~~~~~ \mathbf{L}_{G} \mathbf{u}_i&=\lambda_{i} \mathbf{L}_{P}\mathbf{u}_i,~~i=1,...,n;~~~~~~\\
\textbf{(b)}~~~~\lambda^{}_{max}&=\lambda^{}_{n}\geq \lambda^{}_{n-1}...\geq \lambda^{}_{1}=\lambda^{}_{min};\\
\textbf{(c)  }~~~~~~~\lambda^{(f)}_{1}~~~~~~&\geq \lambda^{(0)}_{1}\overline{\Delta}_{\lambda_{1}}.
\end{split}
\end{equation}


\begin{algorithm}[!htbp]
\small { \caption{Edge Scaling via Constrained SGD Iterations} \label{alg:sgd}
\textbf{Input:} $\mathbf{L}_G$, $\mathbf{L}_P$, $\mathbf{d}_G$, $\mathbf{d}_P$, $\lambda^{(0)}_1$, $\lambda^{(0)}_n$, $\overline{\Delta}_{\lambda_{1}}$, $\alpha$, $\eta_{max}$, $\epsilon$, and $K_{max}$\\
\textbf{Output:} $\tilde{\mathbf{L}}_P$ with scaled edge weights\\
  \algsetup{indent=1em, linenosize=\small} \algsetup{indent=1em}
    \begin{algorithmic}[1]
    \STATE{Initialize: $k=1$, $\eta^{(1)}=\eta_{max}$, $\Delta_{\lambda_1}=\left(\overline{\Delta}_{\lambda_{1}}\right)^{\frac{1}{K_{max}}}$};
     \STATE{For each subgraph edge $(p,q)\in E_P$, initialize $\Delta w^{(1)}_P(p,q)=0$};
     \WHILE {$\left(\mathbf{\frac{\Delta \lambda^{(k)}_n}{\lambda^{(k)}_n}\geq\epsilon}\right)$~$\wedge$~$\left(k\leq K_{max}\right)$}
     	\STATE{Compute approximate eigenvector $\mathbf{h}^{(k)}_t$ by (\ref{formula_pwr_iter5})};
        \FOR{each edge $(p,q)\in E_P$}
        \STATE{$s^{(k)}_{p,q}:=-\lambda^{(k)}_n \left(\mathbf{e}_{p,q}^\top\mathbf{h}^{(k)}_t\right)^2$ by (\ref{formula_weight_sensitivity2})};
        \STATE{$\Delta w^{(k+1)}_P({p,q}):=\alpha \Delta w^{(k)}_P({p,q})-\eta^{(k)} s^{(k)}_{p,q}$ };        
        \STATE{$\phi(p):=\frac{\mathbf{d}_G(p)}{\mathbf{d}_P(p)+\Delta w^{(k+1)}_P({p,q})}$};
        \STATE{$\phi(q):=\frac{\mathbf{d}_G(q)}{\mathbf{d}_P(q)+\Delta w^{(k+1)}_P({p,q})}$};
	\IF{ $\min\left(\phi(p),\phi(q)\right)\leq \lambda^{(k)}_1 \Delta_{\lambda_1}$}
	\STATE{~~~~~~~$\Delta w_p:=\frac{\mathbf{d}_G(p)}{\lambda^{(k)}_1\Delta_{\lambda_1}}-\mathbf{d}_P(p)$};
    \STATE{~~~~~~~$\Delta w_q:=\frac{\mathbf{d}_G(q)}{\lambda^{(k)}_1\Delta_{\lambda_1}}-\mathbf{d}_P(q)$};
    	\STATE{~~~~~$\Delta w^{(k+1)}_P({p,q}):=\min\left(\Delta w_p,\Delta w_q\right)$};
        \ENDIF
        \STATE{$w_P({p,q}):=w_P({p,q})+\Delta w^{(k+1)}_P({p,q})$};
	\STATE{$\mathbf{d}_P(p):=\mathbf{d}_P(p)+\Delta w^{(k+1)}_P({p,q})$};
	\STATE{$\mathbf{d}_P(q):=\mathbf{d}_P(q)+\Delta w^{(k+1)}_P({p,q})$};
        \ENDFOR
        \STATE{$\eta^{(k+1)}:=\frac{\lambda^{(k)}_n}{\lambda^{(0)}_n}\eta_{max}$};
        \STATE{$k:=k+1$};
        \STATE{Update $\lambda^{(k)}_1$ \& $\lambda^{(k)}_n$ by (\ref{formula_courant-fischer-3})};
         \STATE{$\Delta \lambda^{(k)}_n:=\lambda^{(k-1)}_n-\lambda^{(k)}_n$};
     \ENDWHILE
     \STATE {Return the sparsified graph.}
    \end{algorithmic}
    }
\end{algorithm}

In the above formulation, $\lambda^{(0)}_{1}$ and $\lambda^{(f)}_{1}$ represent the smallest nonzero eigenvalue before and after subgraph scaling, respectively. $\overline{\Delta}_{\lambda_{1}}$ represents the upper bound of reduction factor in $\lambda^{(0)}_{1}$ after edge scaling. (\ref{edge_opt}) aims to minimize $\lambda_{n}$ by scaling up subgraph edge weights while limiting the decrease in $\lambda_{1}$. 

A constrained SGD algorithm with momentum \cite{sutskever2013importance} has been proposed  for iteratively scaling up edge weights, as shown in Algorithm \ref{alg:sgd}. The algorithm inputs include: the graph Laplacians $\mathbf{L}_G$ and  $\mathbf{L}_P$,  vectors  $\mathbf{d}_G$ and  $\mathbf{d}_P$  for storing diagonal elements in Laplacians, the initial largest and smallest generalized eigenvalues $\lambda^{(0)}_n$ and $\lambda^{(0)}_1$, the upper bound reduction factor $\overline{\Delta}_{\lambda_{1}}$ for $\lambda_1$, the coefficient $\alpha$ for combining the previous and the latest updates during each SGD iteration with momentum, the maximum step size $\eta_{max}$ for update, as well as the SGD convergence control parameters $\epsilon$ and $K_{max}$. \textbf{Lines 1-2} initialize parameters for the following SGD iterations. \textbf{Line 3} monitors the convergence condition for SGD iterations. \textbf{Lines 6-7} compute the weight update in SGD using the latest sensitivity and the previous update (momentum). \textbf{Lines 8-17} check the impact on $\lambda_1$ due to weight update: if  $\lambda_1$ decreases significantly, an upper bound for weight update is applied; otherwise directly apply the weight update computed in the previous steps.

\subsection{Algorithm Complexity of The Proposed Spectral Graph Reduction Approach}  
\begin{algorithm}[!htbp]
\small { \caption{Algorithm Flow for Spectral Graph Reduction} \label{alg:node_reduction}
\textbf{Input:} Original graph Laplacian $\mathbf{L}_{G}$, user-defined reduction ratio $\psi$, graph density threshold $\gamma_{\textrm{max}}$\\
  \algsetup{indent=1em, linenosize=\small} \algsetup{indent=1em}
    \begin{algorithmic}[1]
    \STATE{Calculate graph density by $\gamma = \frac{|E_G|}{|V|}$};
    \IF { $\gamma < \gamma_{\textrm{max}}$}
    \STATE{Do node reduction (\textbf{Phase A}) on graph $G$ to get graph $R$};
    \STATE{Apply spectral sparsification and edge scaling vis SGD (\textbf{Phase B}) on graph $R$ to get graph $S$};
    \ELSE{}
    \STATE{Apply spectral sparsification and edge scaling vis SGD (\textbf{Phase B}) on graph $G$ to get graph $P$};
    \STATE{Do node reduction (\textbf{Phase A})  on graph $P$ to get graph $S$};
    \ENDIF
     \STATE {Return graph $S$ and $\mathbf{L}_S$}.
    \end{algorithmic}
    }
\end{algorithm} 
 The complete algorithm flow for the proposed spectral graph reduction approach has been shown in Algorithm \ref{alg:node_reduction}. 
 The algorithm complexity of \textbf{Phase (A)} for the spectrum-preserving node reduction procedure  is $O(|E_P|)$ for dense graphs and $O(|E_G|)$ for sparse graphs, the complexity of  \textbf{Phase (B)} for spectral graph sparsification and edge scaling by SGD iterations is $O(|E_G|log(|V|))$ for dense graphs and $O(|E_S|log(|V_R|))$ for sparse graphs. Therefore, the worse-case algorithm complexity of the proposed spectral graph reduction method is $O(|E_G|log(|V|))$. 

\subsection{Solution Refinement by Graph Filters}\label{sect:smoothing}
\subsubsection{Graph Signal Processing and Spectral Sparsification/Reduction}\label{sec:gsp}
 To efficiently analyze signals on undirected graphs, graph signal processing techniques have been extensively studied in recent years \cite{shuman2013emerging}. There are analogies between traditional signal processing or classical Fourier analysis and graph signal processing \cite{shuman2013emerging}: 
 \noindent (1) The signals at different time points in classical Fourier analysis correspond to the signals at different  nodes in an undirected graph;
 \noindent (2) The more slowly oscillating functions in time domain correspond to the graph Laplacian eigenvectors associated with lower eigenvalues or the more slowly varying (smoother) components across the graph. The spectrally sparsified/reduced graphs can be regarded as a ``low-pass" filtered graphs, which have retained as few as possible edges/nodes for preserving the slowly-varying  or ``low-frequency" signals on the original graphs. Consequently, spectrally sparsified/reduced graphs will be able to preserve the eigenvectors associated with low eigenvalues more accurately than high eigenvalues.
 
 \subsubsection{Solution Error due to Spectral Sparsification}\label{sec:gsp}
Denote the ascending eigenvalues and  the corresponding unit-length, mutually-orthogonal eigenvectors  of $\mathbf{L}_G$  by $0 = {\zeta_1} < {\zeta_2} \leq \cdots \leq {\zeta _n}$, and $\mathbf{{\omega _1}, \cdots, {\omega _n}}$, respectively. Similarly denote the eigenvalues and eigenvectors  of $\mathbf{L}_P$ by $0 = {\tilde \zeta _1} < {\tilde \zeta _2} \leq \cdots \leq {\tilde \zeta _n}$ and $\mathbf{{\tilde \omega _1}, \cdots, {\tilde \omega _n}}$, respectively. It should be noted that both $\omega_1$ and $\tilde \omega _1$ are the normalized all-one vector $\mathbf{1}/\sqrt[]{n}$. Then the following spectral decompositions of $\mathbf{L}_G$ and $\mathbf{L}_P$ will hold:
\begin{equation}\label{formula_spec_decomp} 
\mathbf{L}_G = \sum\limits_{i = 1}^{n} {{\zeta _i}\mathbf{{\omega_i}}\mathbf{{\omega^\top_i}}},\mathbf{L}_{P} = \sum\limits_{i = 1}^{n} {{\tilde \zeta _i}\mathbf{\tilde \omega_i}\mathbf{\tilde \omega^\top _i}}. 
\end{equation}

We assume that the  $k$ smallest eigenvalues and their eigenvectors of $\mathbf{L}_{G}$  have been pretty well preserved in $\mathbf{L}_P$, while the remaining $n-k$ higher eigenvalues and eigenvectors are not. Consequently the following approximate spectral decompositions of  $\mathbf{L}_P$ will hold:
\begin{equation}\label{formula_smoothing_P} 
\mathbf{L}_{P} \approx \sum\limits_{i = 1}^{k} {{\zeta _i}\mathbf{\omega_\mathbf{i}}{\omega^\top _i}}+\sum\limits_{i = k+1}^{n} {{\tilde \zeta _i}\mathbf{\tilde \omega_i}\mathbf{\tilde \omega^\top _i}}. 
\end{equation}
In the following, we show that using spectrally-sparsified graphs for solving sparse matrix problems will only result in solution errors that can be expressed with high eigenvectors, while the error analysis for spectrally-reduced graphs will be quite similar and omitted in this work. Consider the following SDD matrix solution problem:
\begin{equation}\label{replace power method G}
(\mathbf{L}_{G}+\delta\mathbf{I})^{}\mathbf{ x}=\mathbf{b^\perp},
\end{equation}
where $\mathbf{b^\perp}\in R^n$ is a random right-hand-side (RHS) vector orthogonal to the all-one vector $\mathbf{1}$, $\delta$ is a very small positive real value added to graph Laplacian for modeling boundary conditions, and $\mathbf{I}\in R^{n\times n}$ is an identity matrix that can be written as follows:
\begin{equation}\label{I}
\mathbf{I}=\sum\limits_{i = 1}^{n}{\mathbf{\omega_i}}{\mathbf{\omega^\top _i}}\approx \sum\limits_{i = 1}^{k} {\mathbf{\omega_\mathbf{i}}{\omega^\top _i}}+\sum\limits_{i = k+1}^{n} {\mathbf{\tilde \omega_i}\mathbf{\tilde \omega^\top _i}},
\end{equation}
we can rewrite $\mathbf{L}_{G}+\delta\mathbf{I}$ as follows:
\begin{equation}\label{replace power method G expand}
\begin{array}{l}
\mathbf{L}_{G}+\delta\mathbf{I}=\sum\limits_{i = 1}^{n} {({\zeta _i+\delta}){\mathbf{\omega_i}}{\mathbf{\omega^\top _i}}}.
\end{array}
\end{equation} 
Consequently,   $\mathbf{x}$ can be written as:
\begin{equation}\label{replace power method G expand2}
\mathbf{ x}=\sum\limits_{i = 1}^{n} \frac{\mathbf{\omega_i\omega^\top _i}\mathbf{b^\perp}}{\zeta _i+\delta}.
\end{equation}
 Let $\tilde{\mathbf{x}}$ denote the approximate solution obtained with $\mathbf{L}_{P}$, then:
\begin{equation}\label{sol G}
\tilde{\mathbf{ x}}\approx  \sum\limits_{i = k+1}^{n} {{{\frac{{\mathbf{\tilde\omega_i}}{\mathbf{\tilde \omega^\top _i}\mathbf{b^\perp}}}{{\tilde \zeta _i+\delta}^{}}}}}+\sum\limits_{i = 1}^{k} {{{\frac{{\mathbf{\omega_i}}{ \mathbf{\omega^\top _i}\mathbf{b^\perp}}}{{ \zeta _i+\delta}}}}},
\end{equation}
which allows us to express the error vector $\mathbf{e}$ as follows:
\begin{equation}\label{err}
\mathbf{e}=\mathbf{ x}-\tilde{\mathbf{x}} \approx\sum\limits_{i = k+1}^{n} \left(\frac{{\mathbf{\omega_i}}{\mathbf{ \omega^\top _i}\mathbf{b^\perp}}}{{ \zeta _i+\delta}}- \frac{{\mathbf{\tilde\omega_i}}{\mathbf{\tilde \omega^\top _i}\mathbf{b^\perp}}}{{\tilde \zeta _i+\delta}}\right).
\end{equation}
(\ref{err}) indicates that when using the sparsified graph Laplacian for solving the SDD matrix, the solution error  can be expressed as a  linear combination  of high eigenvectors corresponding to large Laplacian eigenvalues. Therefore, the error due to the sparsified graph Laplacian  will be a combination of high frequency signals on graphs, which thus can be efficiently filtered out using ``low-pass" graph signal filters \cite{shuman2013emerging}. 
 \subsubsection{Solution Refinement by Smoothing}\label{sec:gsp}
Motivated by recent graph signal processing research \cite{shuman2013emerging}, we   introduce a simple yet effective procedure for improving solutions computed using spectrally sparsified/reduced graphs, which will  enable to  leverage ultra-sparse subgraphs in many numerical and graph algorithms without sacrificing solution quality.  To this end, weighted Jacobi or  Gauss-Seidel methods can be   applied for filtering out such high-frequency error signals on graphs, which have been widely adopted in modern iterative methods for solving large sparse matrices \cite{saad2003iterative}, such as the smoothing (relaxation) function in multigrid algorithms \cite{livne2012lean}. This work adopts a weighted Jacobi iteration scheme for filtering eigenvectors on the graph, while the detailed filtering algorithm has been described in Algorithm \ref{alg:jacobi}. The algorithm inputs include the original Laplacian matrix $\mathbf{L}_G$ that has been decomposed into a diagonal matrix  $\mathbf{D}_G$ and an adjacency matrix   $\mathbf{A}_G$, the approximate solution vectors obtained  using sparsified Laplacian  $\mathbf{L}_P$, as well as the weight $\vartheta$ and iteration number $N_{iter}$ for signal filtering. 
\begin{algorithm}[!htbp]
\small { \caption{Solution Refinement Algorithm} \label{alg:jacobi}
\textbf{Input:} $\mathbf{L}_G=\mathbf{D}_G-\mathbf{A}_G$, $\mathbf{\tilde{x}_1}$,..., $\mathbf{\tilde{x}_{k}}$, $\vartheta$, $N_{iter}$;\\
  \algsetup{indent=1em, linenosize=\small} \algsetup{indent=1em}
\begin{algorithmic}[1]
    \STATE{For each of the approximate solution vectors  $\mathbf{\tilde{x}_1}$,..., $\mathbf{\tilde{x}_k}$, do}
    \FOR{$i=1$ \textbf{to} ${N_{iter}}$ \textbf}
    \STATE{$\mathbf{\tilde{x}^{(i+1)}}=(1-\vartheta)\mathbf{\tilde{x}^{(i)}}+\vartheta\mathbf{D}^{-1}_G\mathbf{A}_G\mathbf{\tilde{x}^{(i)}}$}
    \ENDFOR
     \STATE{Return the solution vectors $\mathbf{\tilde{x}_1}$,..., $\mathbf{\tilde{x}_{k}}$. }
\end{algorithmic}
}
\end{algorithm}

\section{Spectral Reduction for Multilevel Graph Partitioning and Data Visualization}\label{sect:application}
 In this section, multilevel frameworks that allow us to leverage spectrally-reduced graphs for much faster spectral graph partitioning as well as t-distributed Stochastic Neighbor Embedding (t-SNE) of large data sets are introduced. 
\subsection{Multilevel Laplacian Eigensolver for Scalable Spectral Graph Partitioning}


\begin{figure}[htb]
\begin{center}
	\includegraphics[scale=0.42]{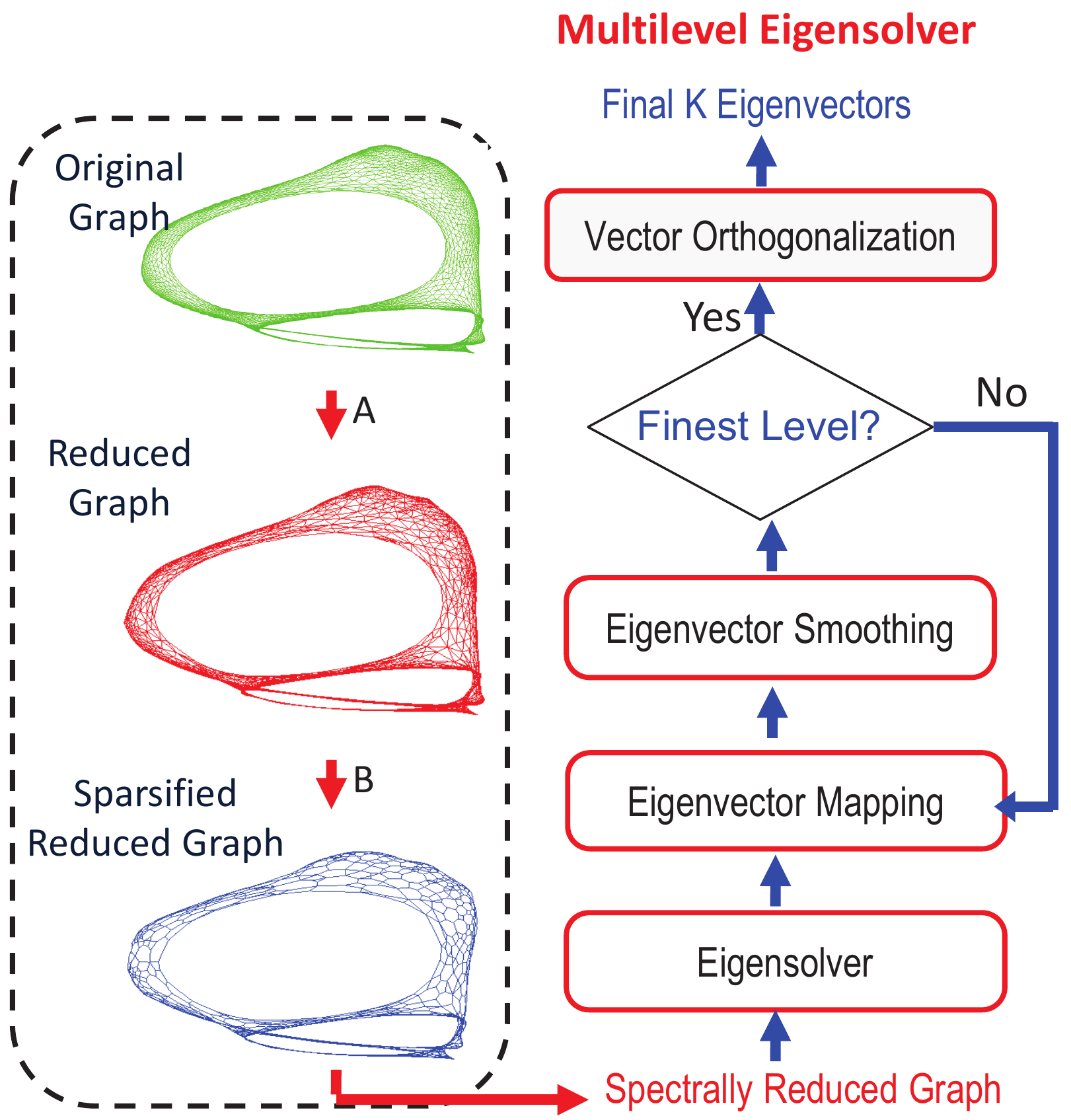}
	\caption{Multilevel Laplacian eigensolver for spectral graph partitioning. \protect\label{fig:eigensolver}}
\end{center}
\end{figure}

\begin{algorithm}[!htbp]
\small { \caption{K-Way Spectral Graph Partitioning} \label{alg:spectralpartiition}
\textbf{Input:} Laplacian matrix $\mathbf{L}_G=\mathbf{D}_G-\mathbf{A}_G$, number of partitions $k$ ;\\
  \algsetup{indent=1em, linenosize=\small} \algsetup{indent=1em}
\begin{algorithmic}[1]
    \STATE{Let $\mathbf{B_G = I}$(ratio cut) or $\mathbf{B_G = D_G}$ (normalized cut)};
    \STATE{Compute the first $k$ eigenvectors $\mathbf{u_1}, \cdots, \mathbf{u_k}$ of eigenproblem $\mathbf{L_Gu_i} = \lambda_i \mathbf{B_Gu_i}$   for $i=1, \cdots, k$};
    \STATE{Form the matrix $\mathbf{U} \in {\rm I\!R^{nxk}}$ with vectors $\mathbf{u_1}, \cdots, \mathbf{u_k}$ as columns};
    \STATE{Cluster the $k$-dimensional points defined by the rows of $\mathbf{U}$ with k-means algorithm};
     \STATE{Return partition $S_1, \cdots, S_k$};
\end{algorithmic}
}
\end{algorithm}
The    k-way spectral graph partitioning (clustering) algorithm has been  described in Algorithm \ref{alg:spectralpartiition} \cite{shi2000normalized, von2007tutorial}, where the  Laplacian eigensolver is usually the computational bottleneck for dealing with large graphs. To this end, we proposed a multilevel Laplacian eigensolver  for more efficiently solving eigenvalue problems by leveraging spectrally-reduced graphs. Note that only the first few nontrivial eigenvectors of the original graph Laplacian are needed for spectral partitioning (clustering) tasks, the spectrally-reduced graphs will thus enable us to solve the eigenvalue problem in a much faster way without loss of solution quality. 

The algorithm flow of the proposed multilevel eigensolver is shown in Figure \ref{fig:eigensolver}. Instead of directly computing the first $k$   eigenvectors of the generalized eigenvalue problem $\mathbf{L_G u_i}=\lambda_i \mathbf{B_G u_i}$, we will first reduce the original graph $G$ into a much smaller graph $S$ such that the eigenvectors of reduced graph can be efficiently calculated.  Next, we will map the eigenvectors of the reduced graph Laplacian onto a finer level using the graph mapping operators (as shown Table \ref{table:symbol}) determined during node aggregation procedure (\textbf{Phase A}). To further improve the approximation quality of these eigenvectors, we apply an eigenvector refinement (smoothing) procedure similar to Algorithm \ref{alg:jacobi}. The eigenvector mapping and smoothing procedures are recursively applied until the finest-level graph is reached. In the last, all eigenvectors for finest-level graph will be orthonormalized through  the Gram-Schmidt process. 

The proposed eigenvector smoothing process is based on the following equations for $i=1, \cdots, k$:
\begin{equation}\label{equ:lgl}
(\mathbf{L_G^{\upsilon}} - \lambda^{\Upsilon}_i \mathbf{B_G^{\upsilon})u^{\upsilon}_i} = 0,
\end{equation}
where $\mathbf{L_G^{\upsilon} = D_G^{\upsilon}-A_G^{\upsilon}}$ is the   Laplacian on level ${\upsilon}$ after   graph reduction, where ${\upsilon}=1$ represents the finest level. We use ${\Upsilon}$ for denoting the coarsest (bottom) level, where $\mathbf{L}_G^{\Upsilon} = \mathbf{L}_S$; $\mathbf{B_G^{\upsilon} = I}$ will be used for ratio cut and $\mathbf{B_G^{\upsilon} = D_G^{\upsilon}}$ for normalized cut (see Section \ref{sec:graph_part} in the Appendix for more details); $\lambda^{\Upsilon}_i$ is the eigenvalue of following generalized eigenproblem:  

\begin{equation}\label{equ:lgs}
\mathbf{L}_G^{\Upsilon}\mathbf{u}^{\Upsilon}_i = \lambda^{\Upsilon}_i \mathbf{B}_G^{\Upsilon}\mathbf{u}^{\Upsilon}_i
\end{equation}

The detailed algorithm for multilevel Laplacian eigensolver is shown in Algorithm \ref{alg:eigensolver}. The inputs of the algorithm include the Laplacian matrix of each hierarchical level $\mathbf L_G^{\upsilon}=\mathbf D_G^{\upsilon}-\mathbf A_G^{\upsilon}$, where $\upsilon = 1, \cdots, \Upsilon$;  mapping operator $\mathbf{H_{{\upsilon}}^{{\upsilon}-1}}$ from level ${\upsilon}$ to level ${\upsilon}-1$ ; and the number of eigenvectors $k$. In the last, spectral partitioning or clustering can be performed using the eigenvectors computed by Algorithm \ref{alg:eigensolver} in the subsequent k-means clustering step. 

\begin{algorithm}[!htbp]
\small { \caption{Multilevel Laplacian Eigensolver} \label{alg:eigensolver}
\textbf{Input:} $\mathbf{L}_G^1, \cdots, \mathbf{L}_G^{\Upsilon}$, \,\, $\mathbf{H}_2^1, \cdots, \mathbf{H}_{{\Upsilon}}^{{\Upsilon}-1}$, \,\, $k$;\\
  \algsetup{indent=1em, linenosize=\small} \algsetup{indent=1em}
    \begin{algorithmic}[1]
    \STATE{Initialize: $j:={\Upsilon}$, $\mathbf{B_G^{\upsilon} := I}$ for ratio cut or $\mathbf{B_G^{\upsilon} := D_G^{\upsilon}}$  for normalized cut, where ${\upsilon} = 1, \cdots, {\Upsilon}$ };
     \STATE{Compute the first $k$ eigenpairs $(\lambda_1^{\Upsilon}, \mathbf{u}_1^{\Upsilon}), \cdots, (\lambda_k^{\Upsilon},\mathbf{u}_k^{\Upsilon})$ of the eigenvalue problem $\mathbf{L}_G^{\Upsilon}\mathbf{u}^{\Upsilon}_i = \lambda_i ^{\Upsilon}\mathbf{B}_G^{\Upsilon}\mathbf{u}^{\Upsilon}_i$ for $i=1, \cdots k$};
      \STATE{Form the matrix $\mathbf{U}^{{\Upsilon}}$ with vectors $\mathbf{u}_1^{\Upsilon}, \cdots, \mathbf{u}_k^{\Upsilon}$ as its columns};
     \WHILE {$ j > 1$}
     	\STATE{Map $\mathbf{U}^j$ from level $j$ to level $j-1$ by $\mathbf{U}^{j-1}= \mathbf{H}_j^{j-1}\mathbf{U}^j$ };
        \FOR{$i=1$ \textbf{to} $k$ }
        \STATE{$\mathbf{y} := \mathbf{U}^{j-1}[:\,,\, i]$, which is the $i$-th column of $\mathbf{U}^{j-1}$}; 
        \STATE{Filter vector $\mathbf{y}$ by performing a few weighted-Jacobi iterations to $(\mathbf{L}^{j-1}_G-\lambda_i^{{\Upsilon}}\mathbf{B}_G^{j-1})\mathbf{y} = 0$} ;
        \STATE{Update $\mathbf{U}^{j-1}[:\, ,\, i]$ with the smoothed vector $\mathbf{y}$} ;
        \ENDFOR
        \STATE{$j:=j-1$};
     \ENDWHILE
     \STATE{Perform orthonormalization to columns of $\mathbf{U}^1$};
     \STATE {Return $\mathbf{U} = \mathbf{U}^1$}.
    \end{algorithmic}
    }
\end{algorithm}

\subsection{Multilevel t-SNE Algorithm for Scalable Data Visualization}

\begin{figure}[htb]
\begin{center}
	\includegraphics[scale=0.465]{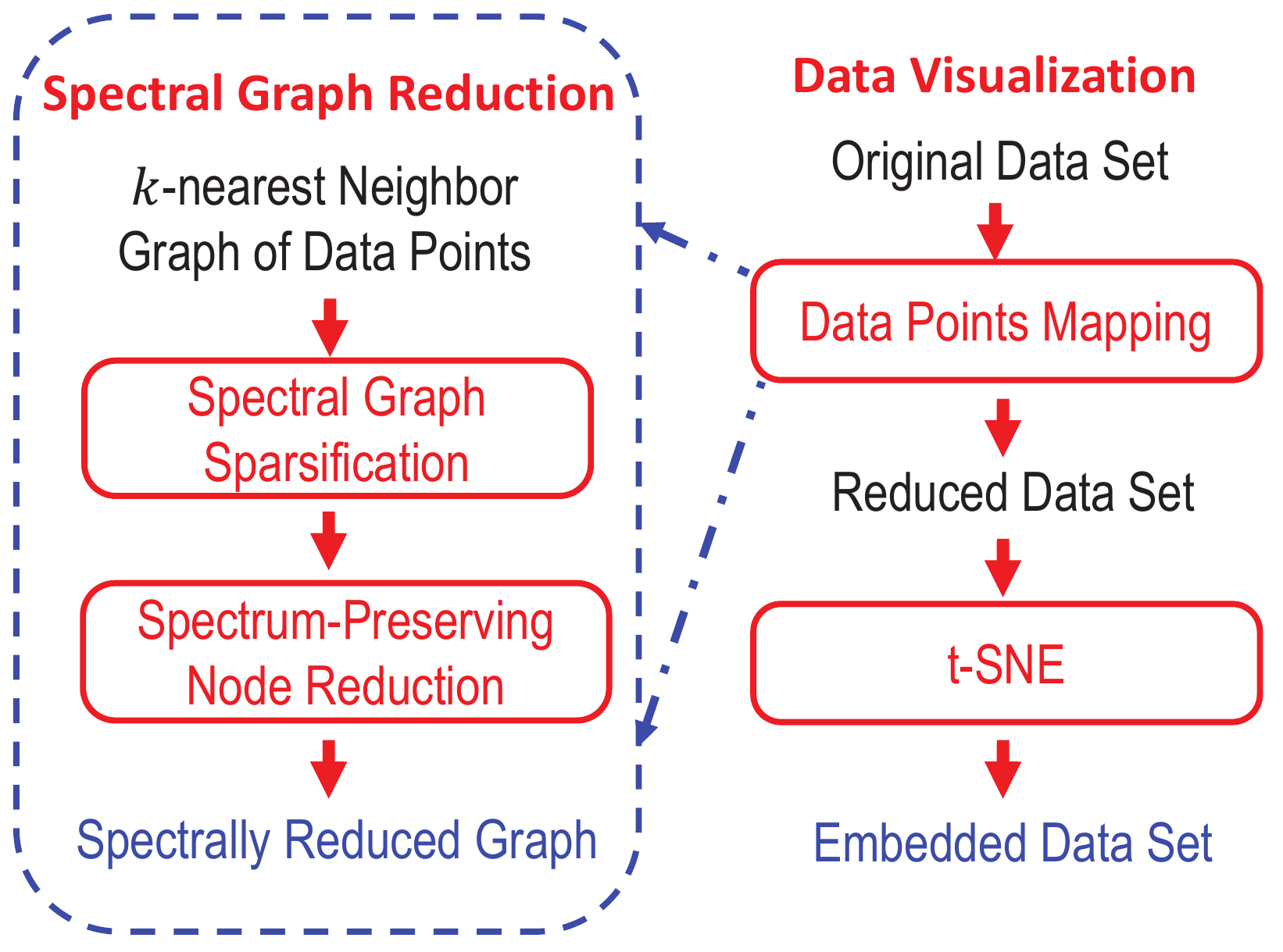}
	\caption{Multilevel t-SNE algorithm. \protect\label{fig:tsne}}
\end{center}
\end{figure}

 Visualization of high-dimensional data is a fundamental problem in data analysis and has been used in many applications, such as medical sciences, physics and economy.  In recent years, the t-Distributed Stochastic Neighbor Embedding (t-SNE) has become the most effective visualization tool due to its capability of performing dimensionality reduction in such a way that the similar data points in high-dimensional space are embedded onto nearby locations in low-dimensional space of two or three dimensions with high probability. However, t-SNE may suffer from very high computational cost for visualizing  large real-world data sets due to the superlinear algorithm computational complexity $O(N^2)$ \cite{maaten2008visualizing, van2014accelerating}, where $N$ is the number of data points in the data set.

Recent research work shows that there is a clear connection between spectral graph partitioning (data clustering) and t-SNE \cite{linderman2017clustering}:  the low-dimensional data points embedding  obtained with t-SNE is closed related to first few  eigenvectors of the corresponding graph Laplacian that encodes the manifold of the original high-dimensional data points. This motivates us to leverage the spectrally-reduced  graphs for computing similar t-SNE embedding results by proposing a multilevel t-SNE algorithm, as described in Algorithm \ref{alg:tsne} and shown in Figure \ref{fig:tsne}. 

The main idea of our multilevel t-SNE algorithm is to aggregate the data points that are closely related to each other on the manifold into much smaller sets, such that visualizing the reduced data set using t-SNE will be much faster and produce similar embedding results. To this end, we start by constructing  a nearest-neighbor (NN) graph, such as the k-NN graph, for the original high-dimensional data points; then a spectrally-reduced (NN) graph is computed using the proposed spectral reduction algorithm. Note that for  k-NN graphs, the graph sparsification and scaling procedure (Phase B) will be  performed before the spectral node aggregation step (Phase A). 

\begin{algorithm}[!htbp]
\small { \caption{Multilevel Data Visualization with t-SNE} \label{alg:tsne}
\textbf{Input:} Original data set $\mathbf{F}$, number of neighbors $k$ ;\\
  \algsetup{indent=1em, linenosize=\small} \algsetup{indent=1em}
\begin{algorithmic}[1]
    \STATE{Generate $k$-nearest neighbor ($k$-NN) graph $G$ based on the data set $F$};
    \STATE{Generate the spectrally-reduced graph $S$};
    \STATE{Form the mapping operators such that $\mathbf{L}_S = \mathbf{H}_G^R\mathbf{L}_G\mathbf{H}_R^G$};
    \STATE{Form a reduced data set $\mathbf{F}_R$ by $\mathbf{F}_R = \mathbf{H}_G^R\mathbf{F}$};
     \STATE{Embed data points with t-SNE on the reduced data set $\mathbf{F}_R$ };
     \STATE{Return embedded data points for visualization}.
\end{algorithmic}
}
\end{algorithm}

\section{Experimental Results}\label{sect:results}

\begin{figure}[htb]
\begin{center}
	\includegraphics[scale=0.34]{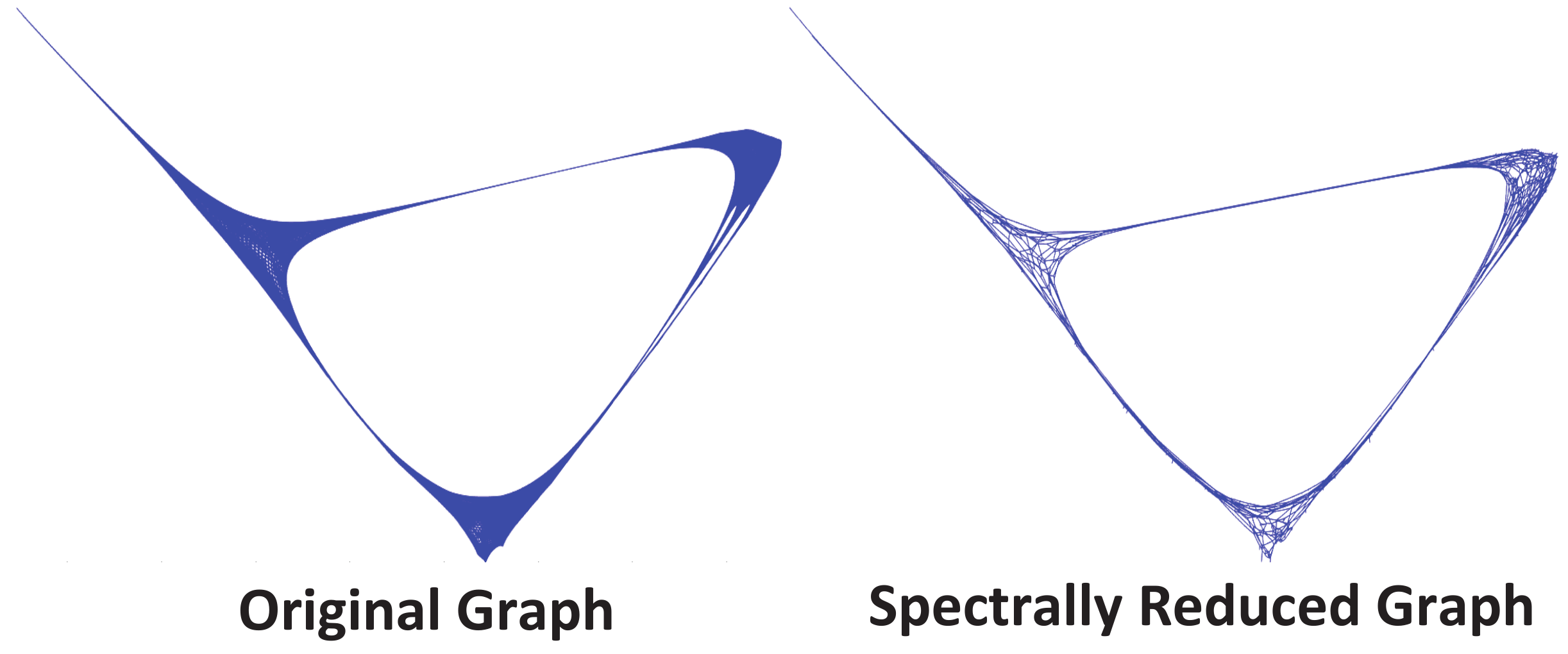}
	\caption{Spectral drawings of the ``fe\_ocean" graph and its reduced graph ($24X$ node reduction and $58X$ edge reduction). \protect\label{fig:oceangraph}}
\end{center}
\end{figure}




\begin{figure}[htb]
\begin{center}
	\includegraphics[scale=0.32]{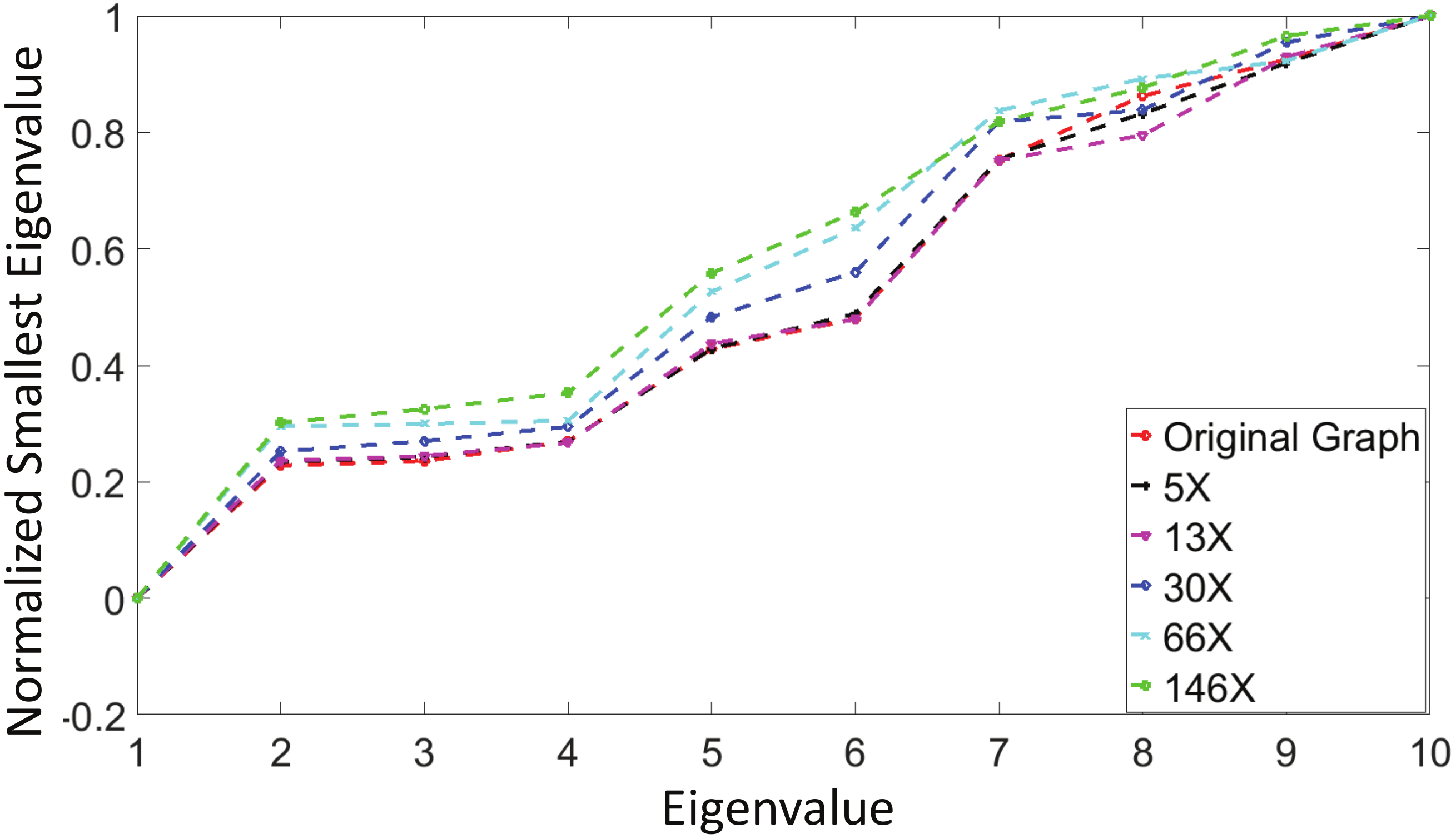}
	\caption{The first 10 normalized eigenvalues of the ``fe\_tooth" graph under different node reduction ratios. \protect\label{fig:eigstooth}}
\end{center}
\end{figure}

In this section, extensive experiments have been conducted to evaluate the proposed spectral graph reduction and spectral partitioning methods with various types of graphs from the DIMACS10 graph collection\cite{bader2014benchmarking, bader2012graph}. Graphs are from different applications,  such as finite-element analysis problems (``fe$\_$tooth", ``fe$\_$rotor") \cite{davis2011matrix}, numerical simulation graphs (``wing\_nodal"), clustering graphs (``uk") and social network graphs (``coAuthorsDBLP" and ``coPapersCiterseer") \cite{davis2011matrix}, etc.  All experiments have been conducted on a single CPU core of a computing platform running 64-bit RHEW 6.0 with 2.67GHz 12-core CPU and 48GB DRAM memory. 

Figure \ref{fig:oceangraph} shows the spectral drawings \cite{koren2003spectral} of the fe$\_$ocean graph and its reduced graph computed by the proposed spectral graph reduction algorithm, where the node and edge reduction ratio are $24X$ and $58X$, respectively. We observe that the spectral drawings of two graphs are highly similar to each other, which indicates very well preserved spectral properties (Laplacian eigenvectors) in the reduced graph. Figure \ref{fig:eigstooth} shows the first few normalized eigenvalues of the original and reduced graph Laplacians, indicating clearly that the smallest eigenvalues of the original Laplacian and the reduced Laplacian match very well even for very large reduction ratios.

Table \ref{testcase} shows spectral graph reduction results on different kinds of graphs using the proposed method, where $T_{reduction}$ denotes the spectral graph reduction time. Compared to other test cases that correspond to sparse graphs, the graph densities of ``${coPapersCiteseer}^\ast$" and ``${appu}^\ast$"  are much higher and thus have been processed as dense graphs.  We want to further emphasize that directly applying the prior algebraic-distance-based node aggregation scheme \cite{chen2011algebraic}  will not produce acceptable results. For example, the node aggregation algorithm failed to generate the reduced graph for ``${appu}^\ast$" due to very high graph density. On the other hand, there will be no issue for dense graphs if we apply \textbf{Phase (B)} for spectral graph sparsification and scaling before the node aggregation phase.



\begin{figure}[htb]
\begin{center}
	\includegraphics[scale=0.55]{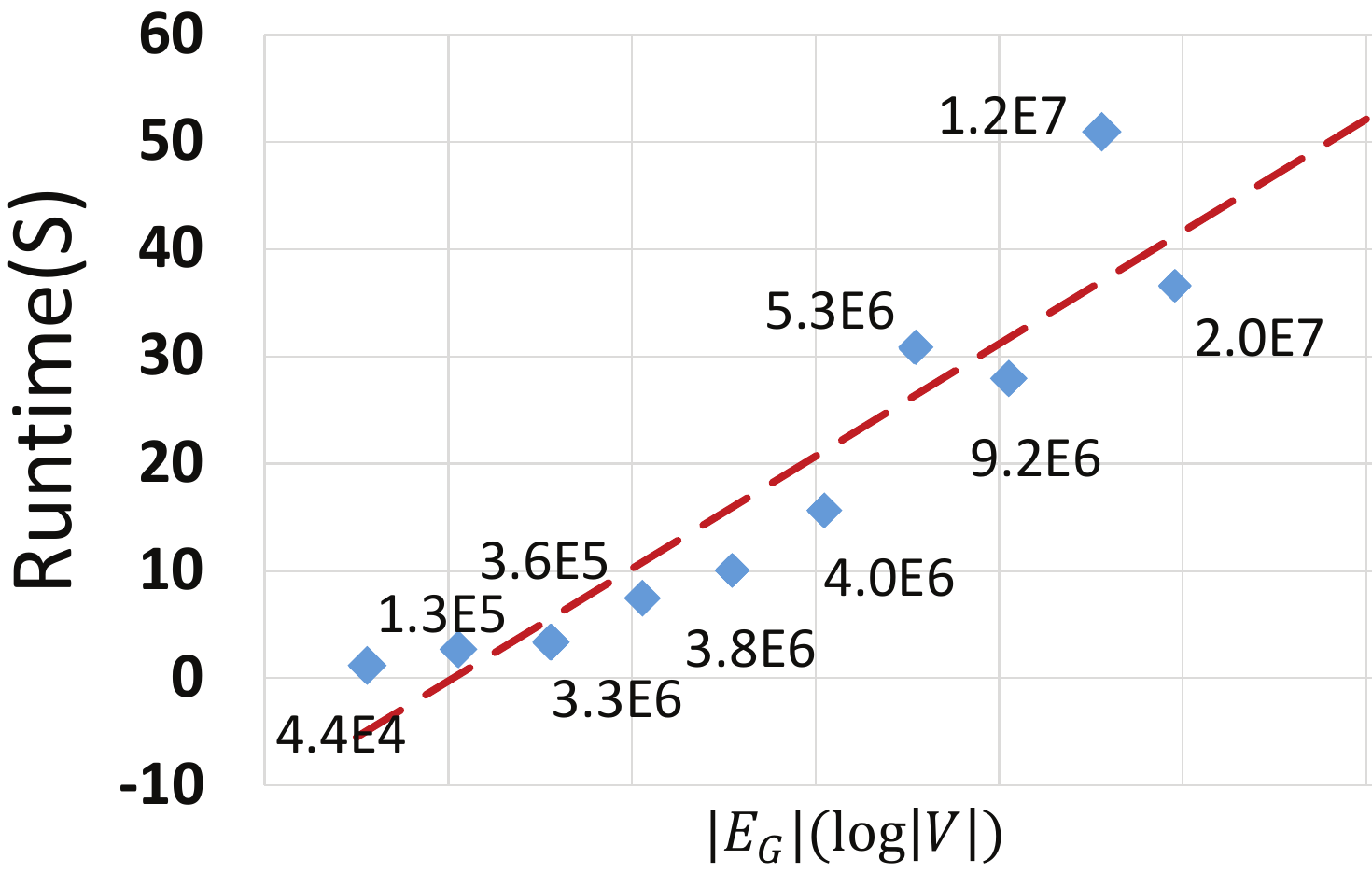}
	\caption{Runtime scalability of proposed spectral graph reduction method. \protect\label{fig:runtime}}
\end{center}
\end{figure}
Figure \ref{fig:runtime} shows the total spectral graph reduction time with different problem sizes $\left(|E_G|\log(|V|)\right)$ for various graphs, where $|E_G|$ ($|V|$) denotes the number of edges (nodes) of the original graphs, respectively. As observed, the total spectral reduction runtime increases almost linearly with the problem size, indicating highly scalable performance of the proposed method ($O\left(|E_G|\log(|V|)\right)$).



\subsection{Results of Scalable Spectral Graph Partitioning}
\begin{table*}
\normalsize
\centering \caption{Spectral Graph Reduction Results on Sample Graphs From DIMACS10 Collection. }
\begin{tabular}{|c|c|c|c|c|c|c|c|}
 \hline
 \multicolumn{3}{|c|}{Test cases }& \multicolumn{2}{|c|}{Original Graph ($G$) }& \multicolumn{3}{|c|}{Spectrally Reduced Graph ($S$ )}\\
 \hline Index & Graph & Application & {$|V|$} & {$|E_G|$} & {$|V_S| \; \left(\frac{|V|}{|V_S|}\right)$} & {$|E_S| \; \left(\frac{|E_G|}{|E_S|}\right)$} & {$T_{reduction}$}\\
 
\hline 1 & fe\_rotor & Finite Element & $1.0E5$ & $6.6E5$  & $1.4E3 \; (71X)$ & $3.7E3 \;(180X)$ & $ 1.30s$\\

\hline 2 & fe\_tooth & Finite Element & $7.8E4$ & $4.5E5$  & $1.3E3 \; (61X)$ & $2.8E3 \;(162X)$ & $ 0.94s$\\


\hline 3 & {auto} & {Numerical simulation} & {$4.5E5$} & {$3.3E6$}  & $1.5E4 \; (30X)$ & $2.0E4 \;(167X)$ & $ 14.81s$\\

\hline 4 & {wing\_nodal} & {Numerical simulation} & {$1.1E4$} & {$7.5E4$}  & $1.8E2 \; (61X)$ & $3.8E2 \;(197X)$ & $ 0.21s$\\

\hline 5 & {luxembourg\_osm} & {Street Network} & {$1.1E5$} & {$1.2E5$}  & $2.6E3 \; (44X)$ & $3.2E3 \;(38X)$ & $ 0.86s$\\

\hline 6 & {mi2010} & {US Census} & {$3.3E5$} & {$7.9E5$}  & $1.3E4 \; (26X)$ & $1.6E4 \;(49X)$ & $ 2.94s$\\

\hline 7 & {uk} & {Clustering} & {$4.8E3$} & {$6.8E3$}  & $1.2E2 \; (40X)$ & $1.3E2 \;(51X)$ & $ 0.22s$\\

\hline 8 & {smallworld} & {Clustering} & {$1.0E5$} & {$5.0E5$}  & $8.2E3 \; (12X)$ & $2.1E4 \;(24X)$ & $ 32.20s$\\

\hline 9 & {vsp\_barth5\_1Kse} & {Star Mixtures} & {$3.2E4$} & {$1.0E5
$}  & {$5.6E2 \; (57X)$} & {$8.3E2 \;(122X)$} & {$0.46s$}\\

\hline 10 & {vsp\_befref\_fxm} & {Star Mixtures} & {$1.4E4$} & {$9.8E4$}  & $2.8E2 \; (49X)$ & $8.1E3 \;(12X)$ & $ 0.24s$\\

\hline 11 & {vsp\_bump2\_e18} & {Star Mixtures} & {$5.6E4$} & {$3.0E5$}  & $3.9E3 \; (14X)$ & $1.3E5 \;(2.3X)$ & $ 0.91s$\\

\hline 12 & {vsp\_p0291\_seymourl} & {Star Mixtures} & {$1.0E4$} & {$5.4E4$}  & $2.0E3 \; (5X)$ & $5.1E3 \;(11X)$ & $ 0.67s$\\

\hline 13 & {vsp\_model1\_crew1} & {Star Mixtures} & {$4.5E4$} & {$1.9E5$}  & $2.1E3 \; (21X)$ & $4.6E3 \;(41X)$ & $ 0.70s$\\

\hline 14 & {vsp\_vibrobox\_scagr7} & {Star Mixtures} & {$7.7E4$} & {$4.4E5$}  & $3.3E3 \; (23X)$ & $9.4E3 \;(47X)$ & $ 2.65s$\\

\hline 15 & {vsp\_bcsstk30\_500sep} & {Star Mixtures} & {$5.8E4$} & {$2.0E6$}  & {$1.7E3 \; (34X)$} & {$3.1E3 \;(654X)$} & {$2.26s$}\\

\hline 16 & {coAuthorsDBLP} & {Citations} & {$3.0E5$} & {$9.8E5$}  & $2.7E4 \; (11X)$ & $3.8E4 \;(26X)$ & $ 30.71s$\\

\hline 17 & {coAuthorsCiteseer} & {Citations} & {$2.2E5$} & {$8.1E5$}  & $2.0E4 \; (11X)$ & $2.5E4 \;(33X)$ & $ 8.20s$\\

\hline 18 & {citationCiteseer} & {Citations} & {$2.6E5$} & {$1.1E6$}  & $2.0E4 \; (13X)$ & $4.1E4 \;(27X)$ & $32.32s$\\


\hline 19 & {coPapersDBLP} & {Citations} & {$5.4E5$} & {$1.5E7$}  & $4.1E4 \; (13X)$ & $7.3E4 \;(210X)$ & $52.83s$\\


\hline 20 & {$coPapersCiteseer^\ast$} & {Citations} & {$4.3E5$} & {$1.6E7$}  & $1.3E4 \; (32X)$ & $1.7E4 \;(950X)$ & $16.41s$\\

\hline 21 & {$appu^\ast$} & {Random Graph} & {$1.4E4$} & {$9.2E5$}  & $2.8E3 \; (5X)$ & $6.7E5 \;(1.4X)$ & $25.53s$\\

 \hline\end{tabular}\label{testcase}
\end{table*}

\begin{table*}
\normalsize
\centering \caption{Results of Graph Partitioning. }
\begin{tabular}{|c|c|c|c|c|c|c|c|c|c|c|}
 \hline
 \multicolumn{2}{|c|}{Test cases} & \multicolumn{3}{|c|}{Original Graph ($G$) } & \multicolumn{4}{|c|}{Spectrally Reduced Graph ($S$ )} & \multicolumn{2}{|c|}{METIS} \\
 
 \hline Index & Graph & $\theta$ & {$T_{eigs}$} & {$T$} & {$\theta$} & {$T_{eigs}$} & {$T_{smooth}$} & {$T$} & $\theta$ & {$T_{metis}$} \\
 
   \hline {1} & fe\_rotor & {$1.51$}  & {$20.2s$} & {$22.8s$} & {$1.50$} & {$0.2s$} & {$2.9s$} & {$5.4s$} & {$1.53$} & {$1.9s$}\\
   
   \hline {2} & fe\_tooth & {$1.77$}  & {$14.6s$} & {$16.6s$} & {$1.68$} & {$0.2s$} &{$1.8s$} & {$4.0s$} & {$1.81$} & {$1.5s$}\\


\hline {3} & {auto} & {$1.10$}  & {$479.7s$} & {$495.8s$} & {$1.08$} & {$0.6s$} & {$12.3s$} &  {$29.0s$} & {$1.16$} & {$3.4s$}\\

\hline {4} & {wing\_nodal} & {$4.88$} & {$2.3s$} & {$3.3s$} & {$4.71$} & {$0.1s$} & {$0.4s$} & {$1.5s$} & {$4.85$}  & {$1.3s$}\\

\hline {5} & {luxembourg\_osm} & {$0.07$} & {$3.5s$} & {$6.3s$} & {$0.07$} &  {$0.2s$} & {$0.9s$} & {$3.8s$} & {$0.07$} & {$4.8s$}\\

\hline {6} & {mi2010} & {$0.43$} & {$14.5s$} & {$21.6s$} & {$0.41$} &  {$0.4s$} & {$3.7s$} &  {$10.2s$} & {$0.49$}  & {$2.6s$}\\

\hline {7} & {uk} & {$1.03$} & {$0.2s$} & {$0.6s$} & {$1.05$}  & {$0.1s$} & {$0.1s$} & {$0.6s$} & {$1.29$}  & {$1.1s$}\\


\hline {8} & {smallworld} & {$7.02$} & {$16,137.9s$} & {$16,144.5s$} & {$7.05$} & {$9.2s$} & {$2.8s$} & {$14.1s$} & {$7.50$} & {$3.1s$}\\
 
\hline {9} & vsp\_barth5\_1Kse & {$3.12$} & {$14.4s$} & {$16.6s$} & {$2.72$} &  {$0.2s$} & {$0.5s$} & {$2.7s$} & {$3.50$} & {$1.2s$}\\

\hline {10} & {vsp\_befref\_fxm} & {$13.59$} & {$3.4s$} & {$4.7s$} & {$12.83$} & {$0.1s$} & {$0.4s$} & {$1.8s$} & {$20.01$} & {$1.8s$}\\

\hline {11} & {vsp\_bump2\_e18} & {$14.60$}  & {$123.0s$} & {$124.7s$} & {$13.55$} & {$1.7s$} & {$1.4s$} & {$5.4s$} & {$16.64$} & {$3.3s$}\\

\hline {12} & {vsp\_p0291\_seymourl} & {$8.09$}  & {$2.2s$} & {$2.9s$} & {$7.88$} & {$0.4s$} & {$0.2s$} & {$1.3s$} & {$16.13$} & {$1.8s$}\\

\hline {13} & {vsp\_model1\_crew1} & {$11.38$}  & {$11.5s$} & {$13.9s$} & {$10.48$} & {$0.7s$} & {$0.8s$} & {$4.9s$} & {$16.94$}  & {$2.4s$}\\

\hline {14} & {vsp\_vibrobox\_scagr7} & {$6.92$}  & {$73.8s$} & {$75.8s$} & {$6.85$} & {$0.6s$} & {$2.3s$} & {$4.8s$} & {$11.75$} & {$2.7s$}\\
 
 \hline {15} & vsp\_bcsstk30\_500sep & {$\dagger$} & {$\dagger$} & {$\dagger$} & {$2.09$} & {$0.2s$} & {$24.0s$} &  {$25.7s$} & {$25.04$} & {$8.3s$}\\
 
\hline {16} & {coAuthorsDBLP} & {$0.92$}  & {$245.3s$} & {$250.8s$} & {$0.49$} &  {$15.7s$} & {$4.2s$} & {$26.5s$} & {$4.75$}  & {$4.8s$}\\

\hline {17} & {coAuthorsCiteseer} & {$0.67$} & {$77.0s$} & {$81.3s$} & {$0.41$} &  {$5.4s$} & {$3.2s$} & {$13.3s$} & {$3.02$}  & {$3.2s$}\\

\hline {18} & {citationCiteseer} & {$0.48$} & {$2,005.2s$} & {$2,027.7s$} & {$0.52$}  & {$12.9s$} & {$4.9s$} & {$24.8s$} & {$5.16$} & {$5.2s$}\\


\hline {19} & {coPapersDBLP} & {$NA$} & {$NA$}& {$NA$} & {$0.14$}  & {$17.4s$} & {$43.1s$} & {$61.6s$} & {$4.01$}  & {$7.8s$}\\


\hline {20} & {coPapersCiteseer*} & {$NA$}  & {$NA$} & {$NA$} & {$0.06$} & {$0.87s$} & {$44.0s$} & {$51.6s$} & {$2.33$} & {$5.8s$}\\

\hline {21} & {appu*} & {$22.47$}  & {$178.9s$} & {$179.9s$} & {$23.80$} & {$7.3s$} & {$3.4s$} & {$11.7s$} & {$27.54$} & {$3.5s$}\\
 \hline\end{tabular}\label{evaluation}
\end{table*}


\begin{figure}[htb]
\begin{center}
	\includegraphics[scale=0.45]{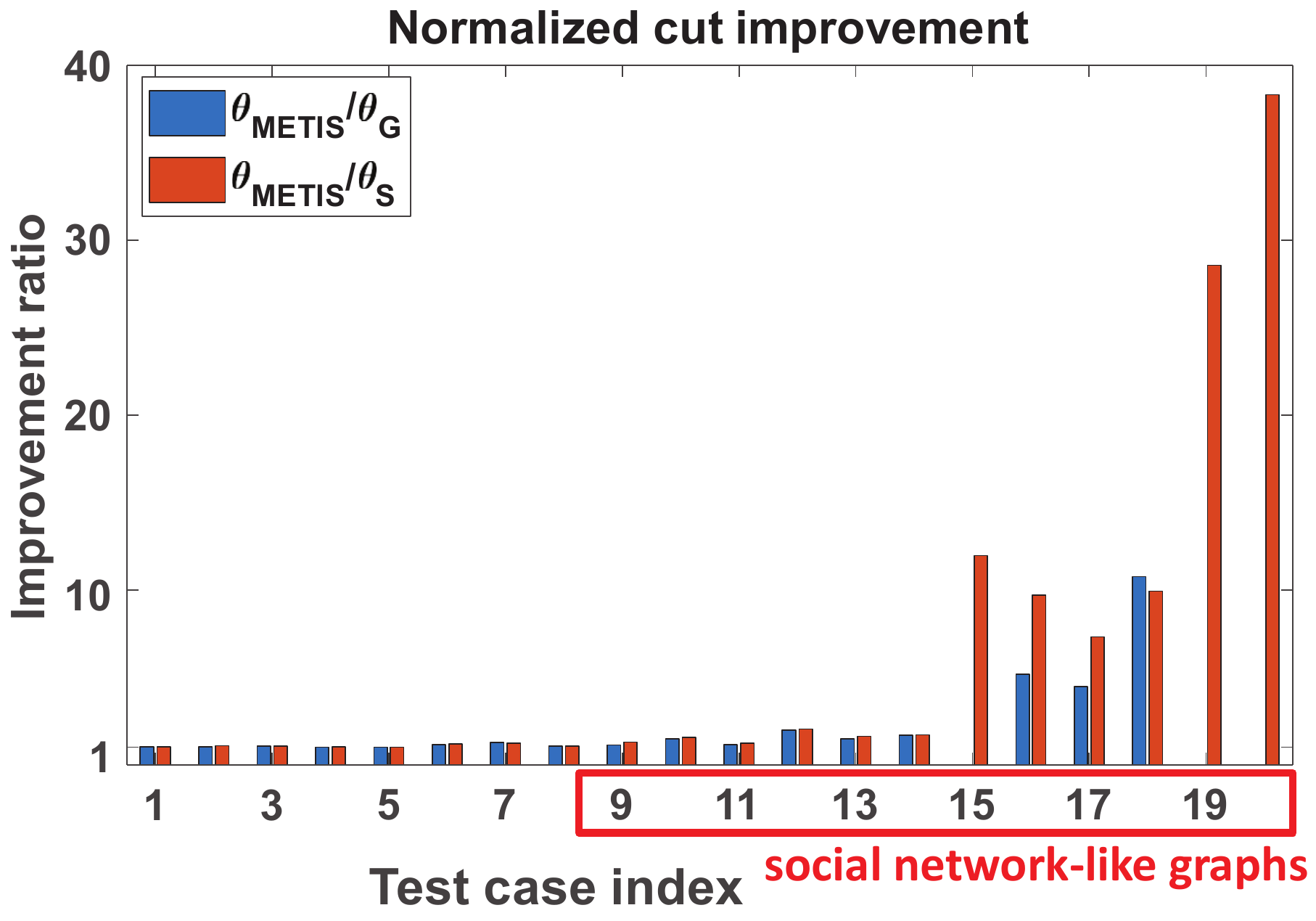}
	\caption{Normalized cut (partitioning quality) improvements over METIS when using spectral partitioning with the original graphs and reduced graphs. \protect\label{fig:theta}}
\end{center}
\end{figure}


\begin{figure}[htb]
\begin{center}
	\includegraphics[scale=0.52]{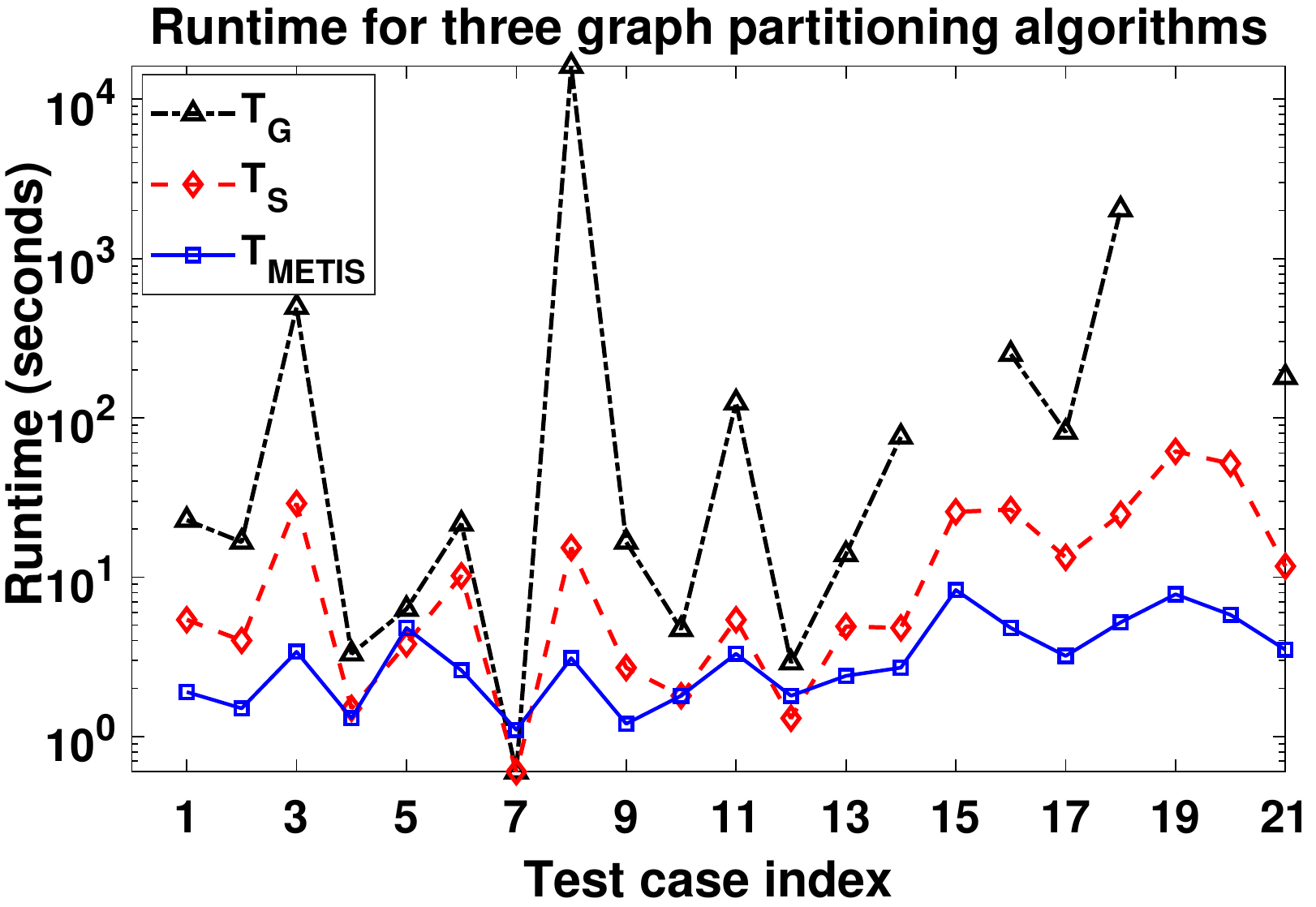}
	\caption{Execution time for graph partitioning when using the original graphs, spectrally reduced graphs and METIS.\protect\label{fig:partitionruntime}}
\end{center}
\end{figure}


\begin{figure}[htb]
\begin{center}
	\includegraphics[scale=0.340]{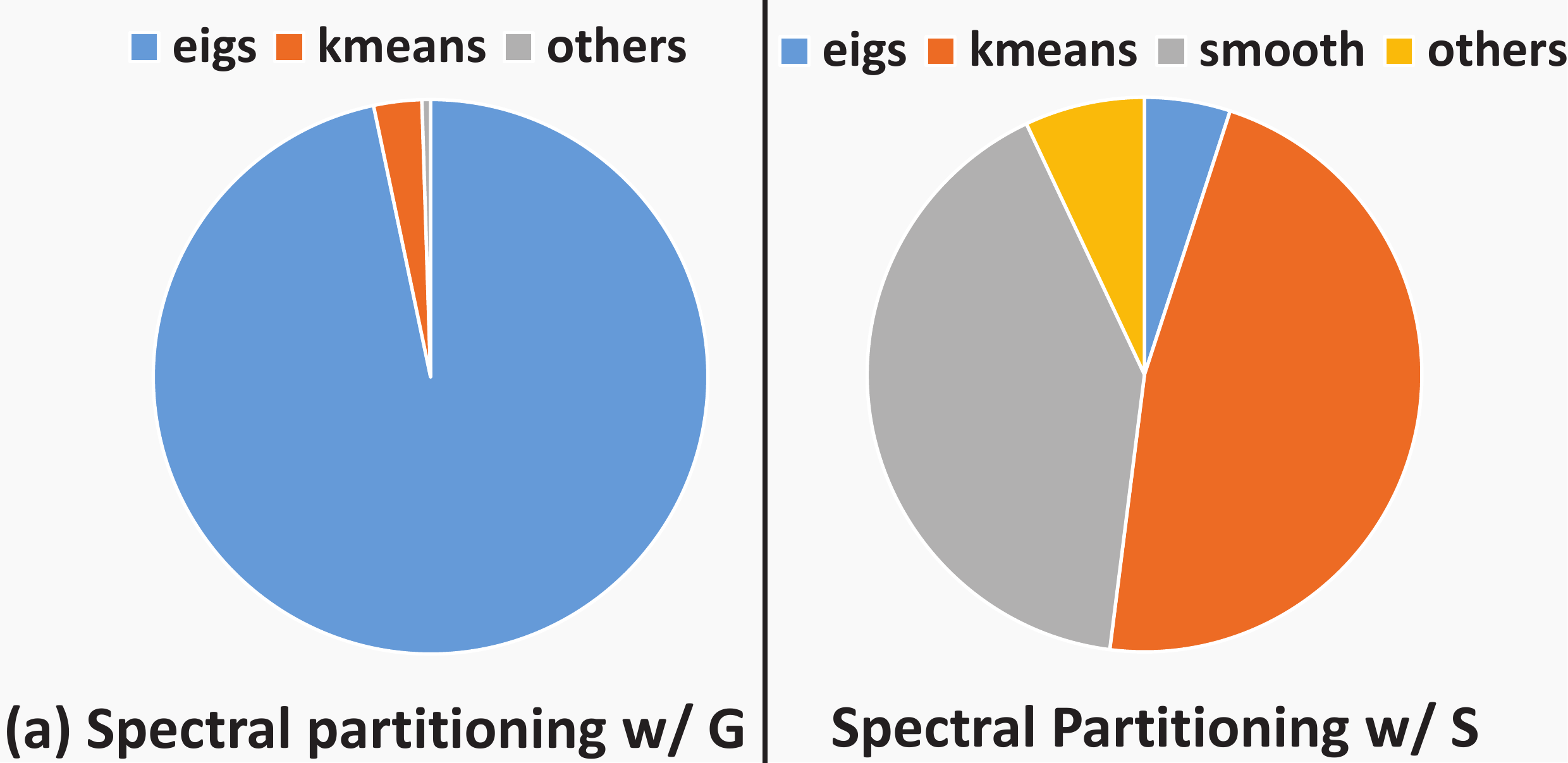}
	\caption{Profiling of time spent in spectral partitioning on ``auto" graph \cite{davis2011matrix}. \protect\label{fig:pie}}
\end{center}
\end{figure}


\begin{figure}[htb]
\begin{center}
	\includegraphics[scale=0.43]{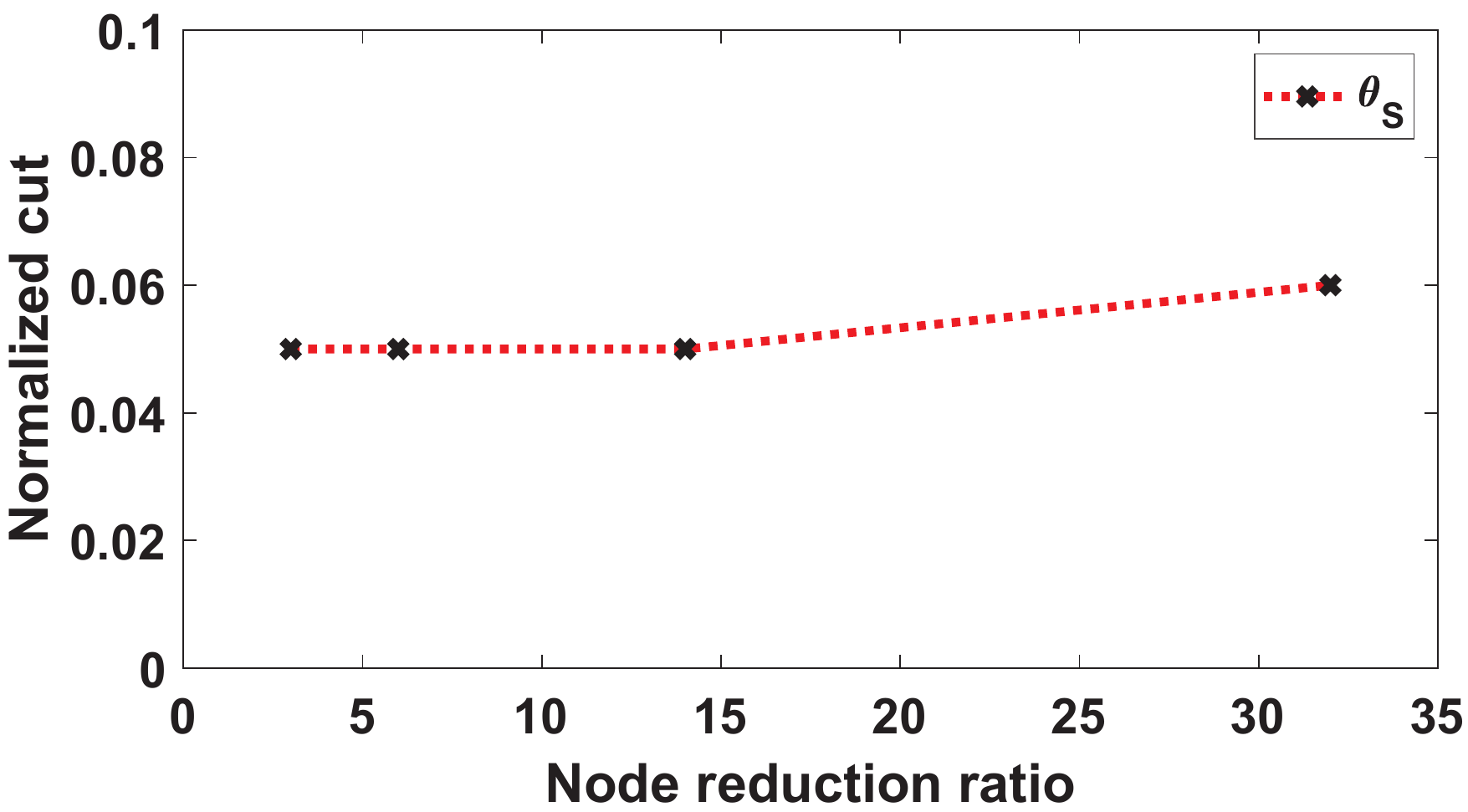}
	\caption{Partitioning qualities (normalized cut) under different reduction ratio for the ``coPapersCiteseer" graph \cite{davis2011matrix}.  \protect\label{fig:ratiotheta}}
\end{center}
\end{figure}


\begin{figure}[htb]
\begin{center}
	\includegraphics[scale=0.580]{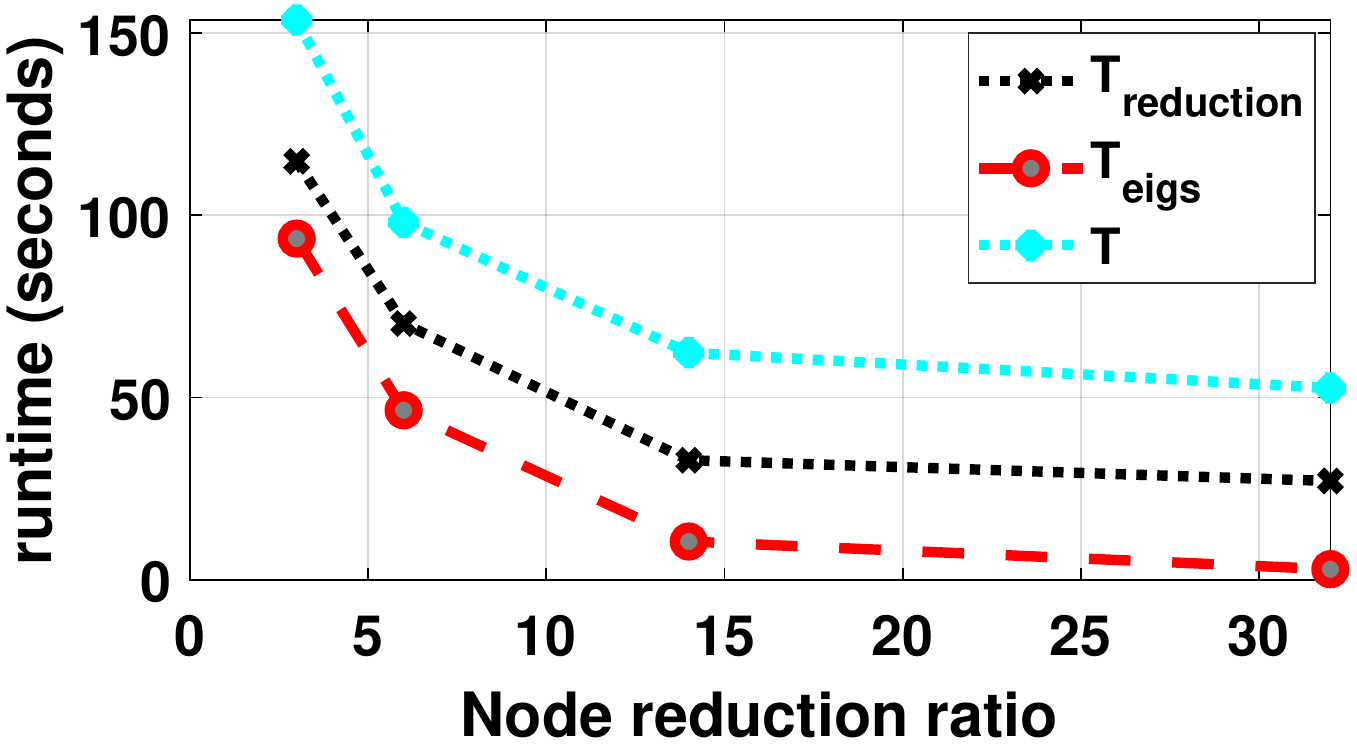}
	\caption{Runtime for multi-way spectral partitioning under different reduction ratio for the ``coPapersCiteseer" graph \cite{davis2011matrix}.  \protect\label{fig:ratioruntime}}
\end{center}
\end{figure}


\begin{figure}[htb]
\begin{center}
	\includegraphics[scale=0.45]{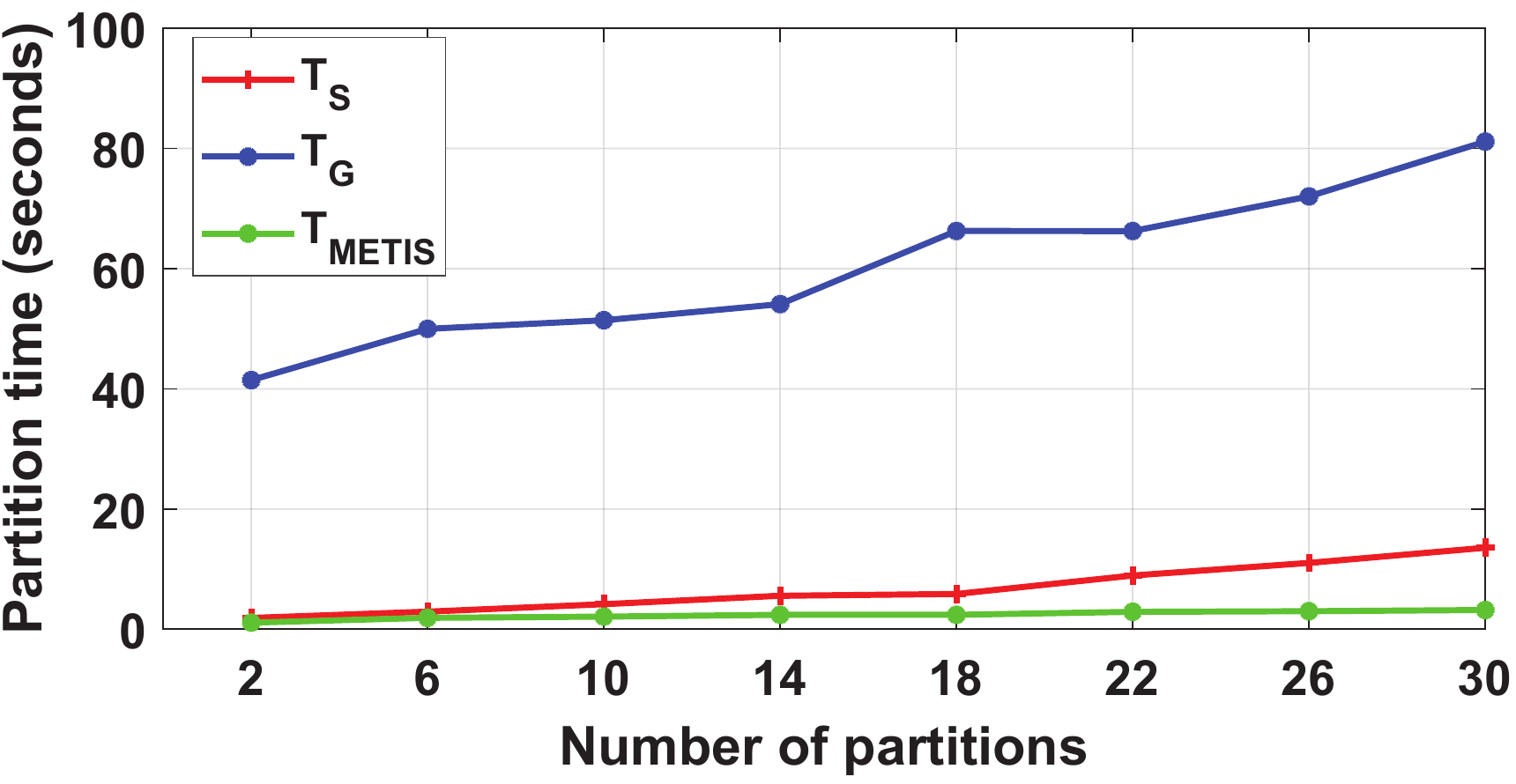}
	\caption{Runtime for graph partitioning with different partitions for the ``coAuthorsCiteseer" graph \cite{davis2011matrix}.  \protect\label{fig:multik_time}}
\end{center}
\end{figure}


\begin{figure}[htb]
\begin{center}
	\includegraphics[scale=0.45]{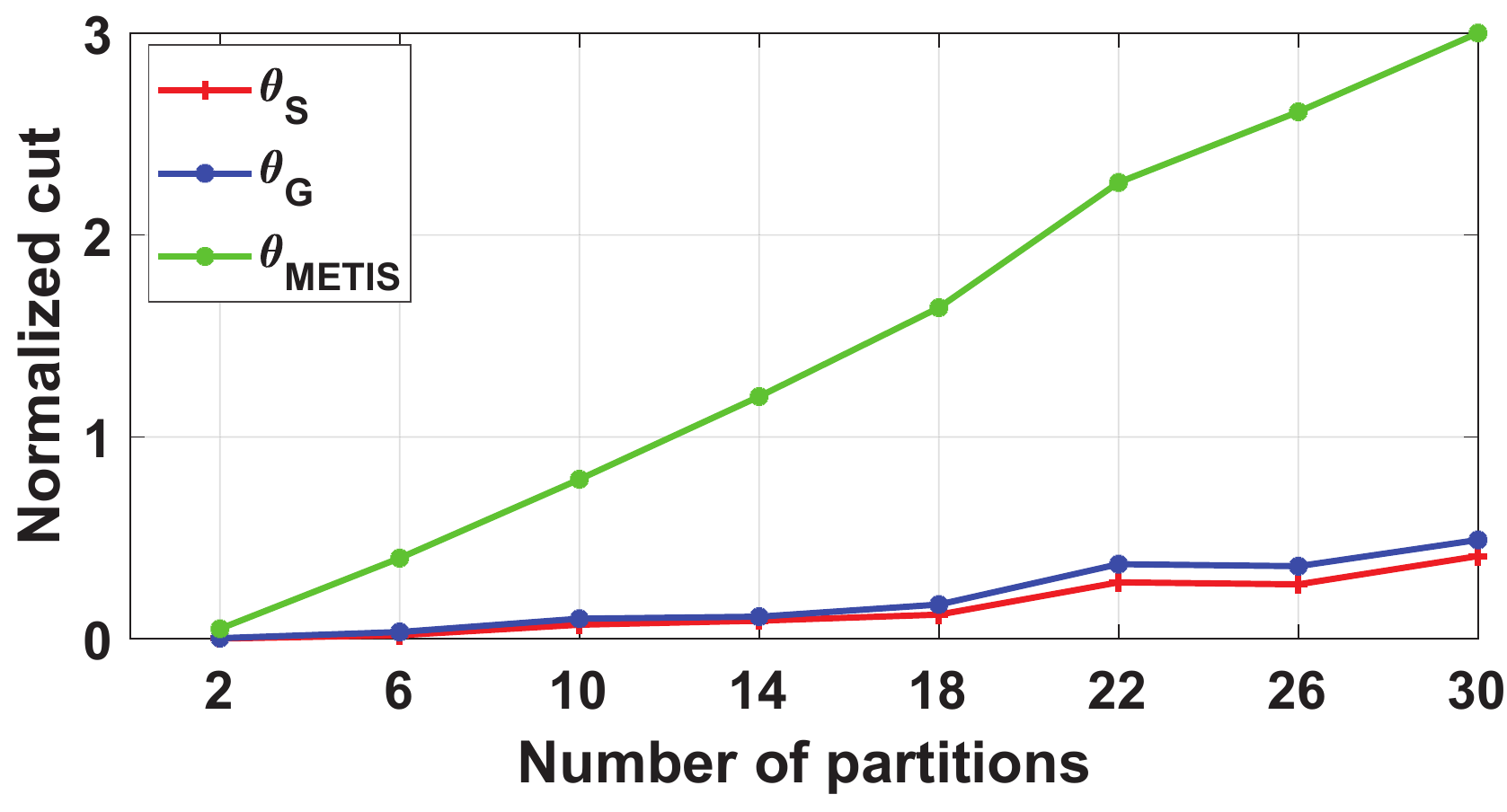}
	\caption{Normalized cut for graph partitioning with different partitions for the ``coAuthorsCiteseer" graph \cite{davis2011matrix}.  \protect\label{fig:mulitk_theta}}
\end{center}
\end{figure}

We evaluated the performance of proposed spectral graph partitioning algorithm on varieties of graphs from the DIMACS10 graph collection. We choose to partition all the graphs into $30$ partitions. The built-in ``eigs" and ``kmeans" MATLAB functions are used for solving the eigenvalue problem and node clustering tasks, respectively. The normalized cut  is used to measure the quality of partitions. Even though the ratio cut and normalized cut are similar, they are trying to solve slightly different optimization problems and one might be preferable over the other depending on the application. Three partitioning algorithms have been tested, including spectral partitioning with original graphs (no reduction), spectral partitioning with graph reduction and the state-of-the-art graph partitioning tool METIS \cite{karypis1995metis}. We use METIS version 5.1.0 and the default parameters in our experiments. The performance of partitioning is evaluated based on the normalized cut and total execution time. Detailed results have been shown in Table \ref{evaluation}, where $\theta$ is the normalized cut, $T_{eigs}$ is the execution time for solving the eigenvalue problem, $T_{smooth}$ denotes eigenvector refinement (smoothing) time, $T$ denotes the total runtime for spectral graph partitioning, $T_{metis}$ is the total time for graph partitioning using METIS, $\dagger$ represents the failure of solving eigenvalue problems due to the singularity of the Laplacian matrix,  and ``NA" denotes the failure of solving eigenvalue problems due to the limited memory resources. From the table we can observe that the overall quality of generated partitions by spectral partitioning is better than the partitioning quality generated by METIS, even though METIS is usually much faster than spectral partitioning. To better compare the performance of the three algorithms, we plot the partitioning quality improvements (ratios of normalized cut) over METIS when using spectral partitioning with the original graph and the reduced graph in Figure \ref{fig:theta}. Meanwhile, the total execution times required by three graph partitioning algorithms have also been shown in Figure \ref{fig:partitionruntime}. Based on these results, the following observations can be made:\\

(1) For sparse graphs with relatively low node degrees, the partition qualities obtained using spectral partitioning and METIS algorithms are comparable;

(2) The spectral partitioning methods can obtain significantly better partitioning results for scale-free graphs, such as the social-network-like graphs that have highly varying node degrees. For example, spectral partitioning algorithm can achieve $39X$ higher partitioning quality than METIS on the ``coPapersCiteseer" graph when considering the normalized cut metric. 

Since METIS is a multilevel graph partitioning algorithm  that relies on  local graph information for constructing the hierarchical levels, coarse level graphs can well preserve the structure of original graph for regular low degree graphs, but not for social-network-like graphs with highly  irregular node degrees \cite{naumov2016parallel}. On the other hand, the proposed multilevel spectral partitioning  scheme will preserve the key spectral properties on spectrally-reduced graphs, making it more reliable for partitioning general graphs. 

Compared to the spectral partitioning method with original graphs, the runtime of spectral partitioning using reduced graph is generally much less, especially for large graphs. For example, we  achieve over $1100X$ runtime speedup on the ``smallworld" graph. For larger graphs, such as the ``coPapersDBLP" and ``coPapersCiteseer" graphs, spectral partitioning without  reduction will fail due to the extremely high computation (memory) cost for running MATLAB's built-in eigensolver. 

Figure \ref{fig:pie} shows the profiling of time required in spectral partitioning of the ``auto" graph. It indicates that most of the runtime is due to the eigensolver if the original graph is used, while the k-means and smoothing time will be dominant when using the spectrally-reduced graph. However, the smoothing procedure is inherently highly parallel making it possible to further improve the efficiency of the proposed spectral partitioning and to develop high-quality parallel spectral partitioning algorithms.

We also evaluated the performance of the proposed spectral partitioning method using different reduction ratios, as shown in Figure \ref{fig:ratiotheta} and Figure \ref{fig:ratioruntime}. We observe that higher graph reduction ratios immediately result in lower cost for graph reduction as well as  spectral partitioning, while still maintaining high partitioning quality. This indicates very promising performance in efficiency and reliability achieved by the proposed algorithm. 

Finally, we evaluate the performance of three partitioning algorithms using different numbers of partitions. As shown in Figure \ref{fig:multik_time} and Figure \ref{fig:mulitk_theta}, the reduced graph has $11\times$ fewer nodes and $26\times$ fewer edges compare to the original graph. With the increasing number of partitions, we observed that  the spectral  partitioning method using the spectrally-reduced graph is slightly slower than METIS but consistently produces higher partitioning qualities. 


\subsection{Results of Scalable Data Visualization}
\begin{figure}
\centering \epsfig{file=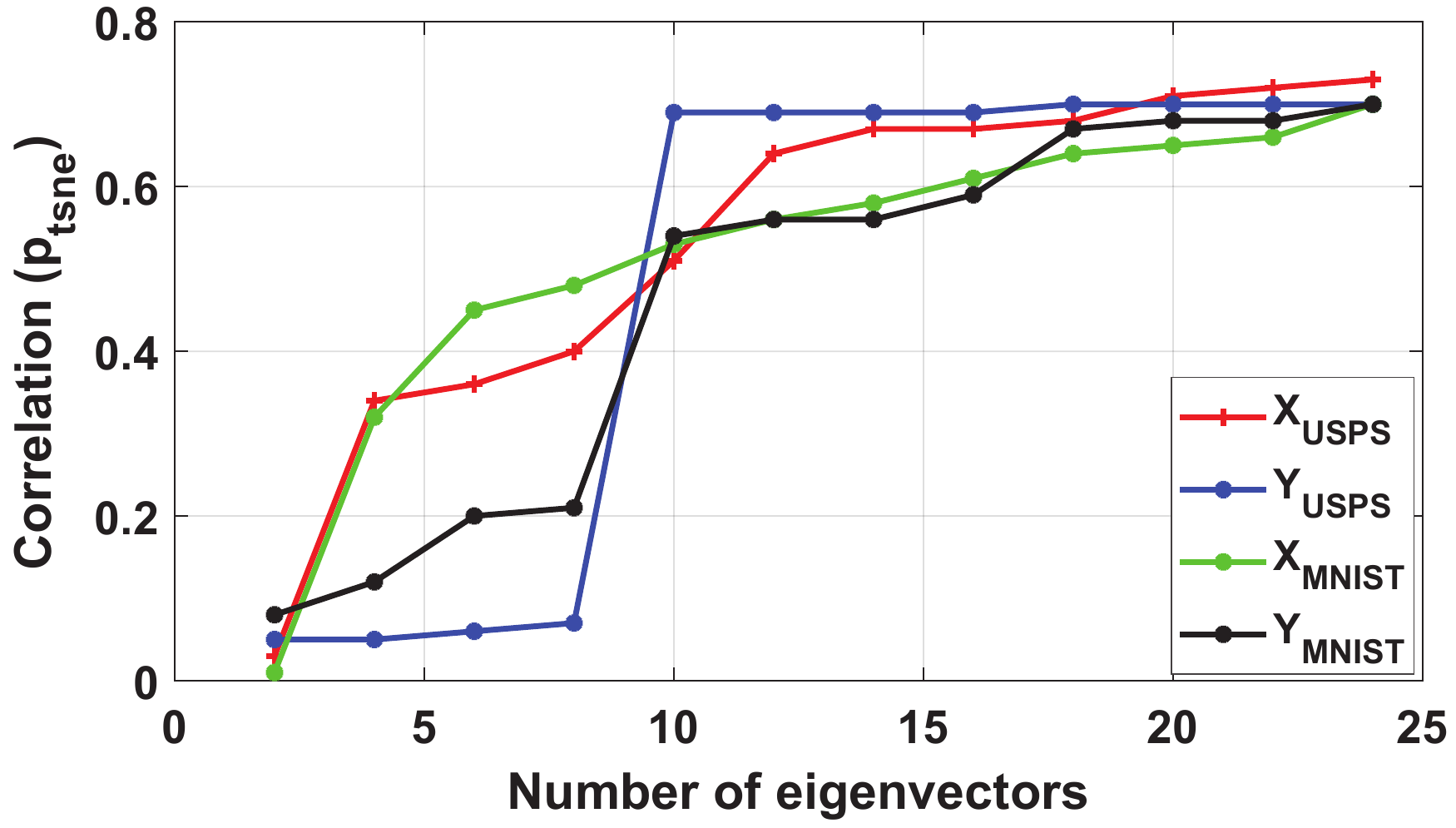, scale=0.49} \caption{Correlations ($X_{USPS}$ and $X_{MNIST}$ for $p^x_{tsne}$; $Y_{USPS}$ and $Y_{MNIST}$ for $p^y_{tsne}$) between   2D embedding vectors computed by t-SNE and the subspace formed by the first few eigenvectors of the Laplacian matrices computed using USPS and MNIST data sets. \protect\label{fig:correlation}}
\end{figure}

We first demonstrate the connection between the t-SNE embedding solution and the first few unnormalized Laplacian eigenvectors of the k-NN graph formed using the original data set. To quantitatively estimate their correlations, we increase the number of Laplacian eigenvectors for representing the embedding vectors $\mathbf{x} \in \mathbb{R}^n$ and $\mathbf{y} \in \mathbb{R}^n$ that store the locations of $n$ data points in 2D space  obtained by running t-SNE, and compute the correlation factors $p^x_{tsne}=\frac{\mathbf{||UU^\top x||_2}}{\mathbf{||x||_2}}$ and $p^y_{tsne}=\frac{\mathbf{||UU^\top y||_2}}{\mathbf{||y||_2}}$, where $\mathbf{U}\in \mathbb{R}^{n\times r}$ is the matrix with the first $r$ Laplacian eigenvectors (of the original k-NN graph) as its column vectors. If $p^x_{tsne}$ or  $p^y_{tsne}$ is close to $1$, it indicates  a strong correlation (significant overlap) between the eigenspace formed by the first few Laplacian eigenvectors and the t-SNE embedding vectors. Figure \ref{fig:correlation} shows   strong correlations   between the low-dimensional embedding vectors generated by t-SNE and the first few (e.g. $r=20$) eigenvectors of the Laplacian matrices corresponding to the k-NN graphs constructed  using the USPS and MNIST data sets \footnote{\textbf{USPS} includes   $9,298$ images of USPS hand written digits with $256$ attributes;
\textbf{MNIST} is a data set from Yann LeCun's website  http://yann.lecun.com/exdb/mnist/, which includes  $70,000$ images of hand written digits with each of them represented by $784$ attributes.}. It is also interesting to observe that the t-SNE embedding vectors are more closely related to the $10$-th eigenvector, since the inclusion of such an eigenvector leads to significantly improved correlation factors $p^x_{tsne}$ and $p^y_{tsne}$. This is actually very reasonable  considering the ground-truth number of clusters for the USPS and MNIST data sets is $10$.

\begin{figure}[htb]
\begin{center}
	\includegraphics[scale=0.55]{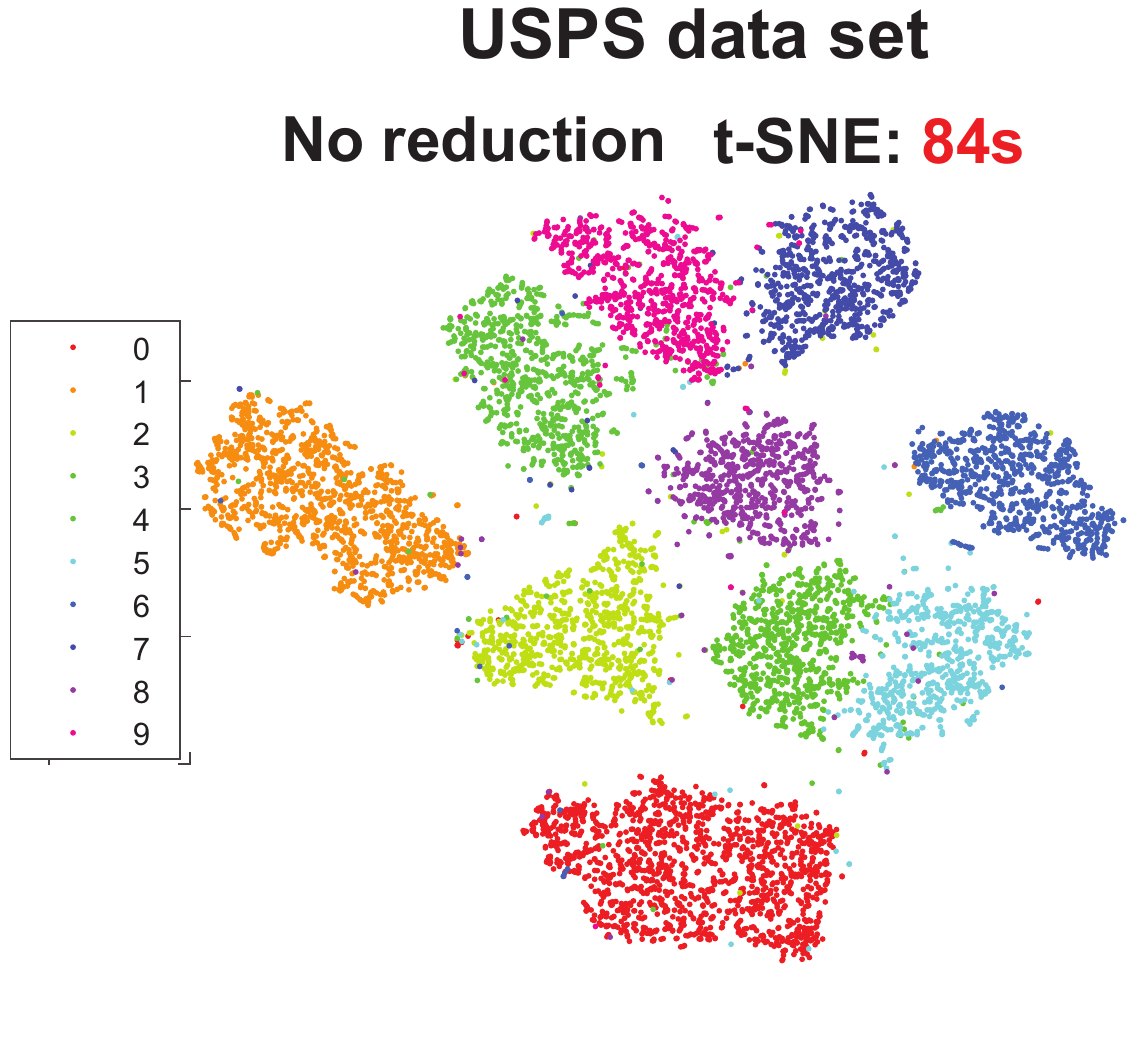}
	\includegraphics[scale=0.55]{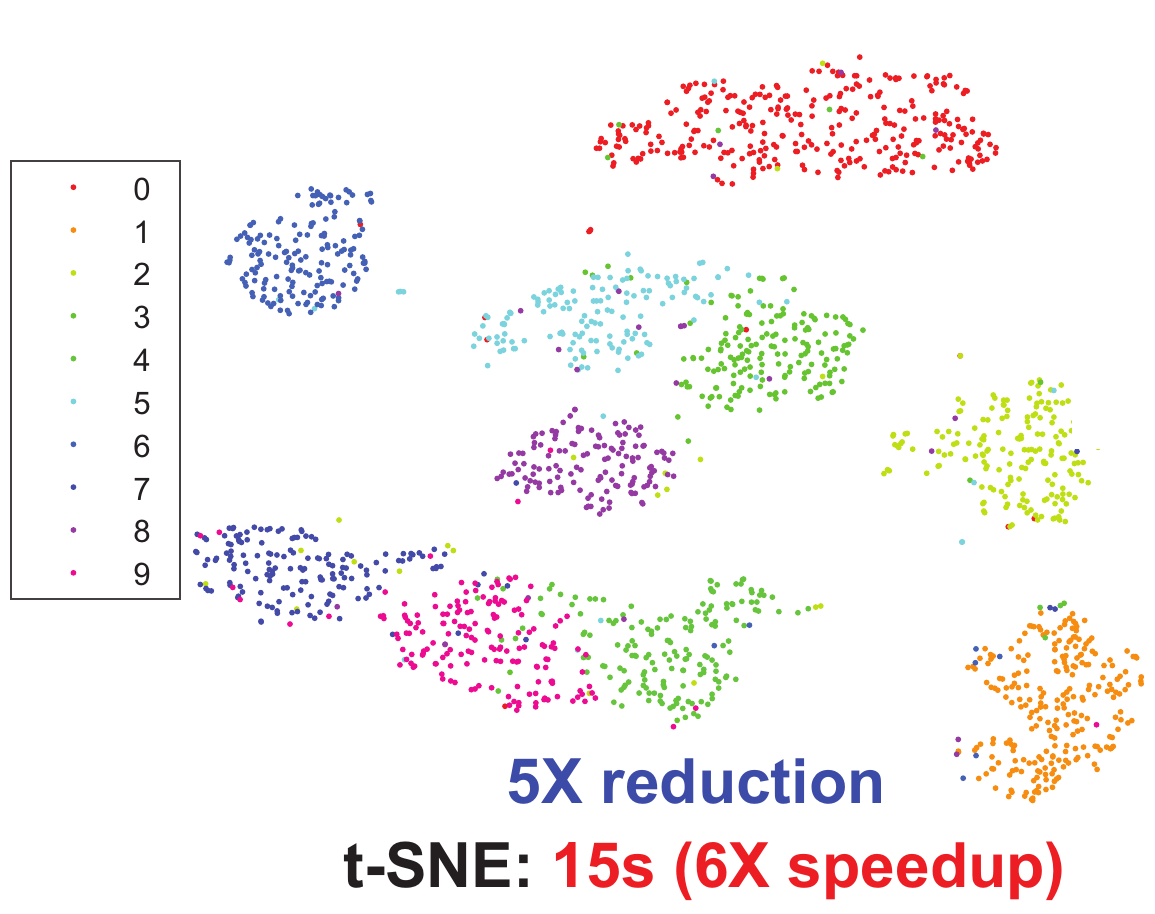}
	\caption{t-SNE visualization with original USPS data set and the reduced data set. \protect\label{fig:usps}}
\end{center}
\end{figure}

\begin{figure}[htb]
\begin{center}
	\includegraphics[scale=0.38]{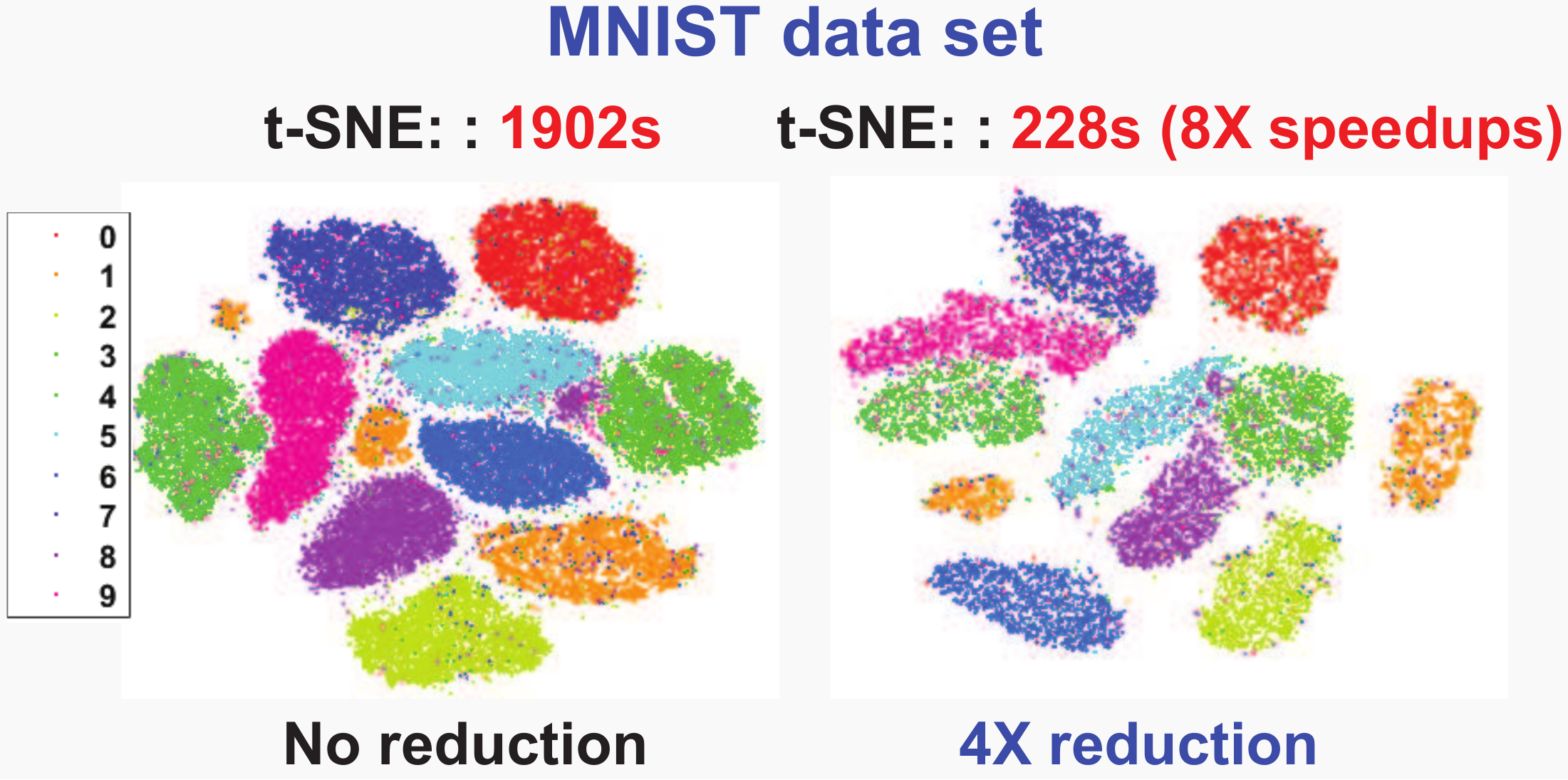}
	\includegraphics[scale=0.38]{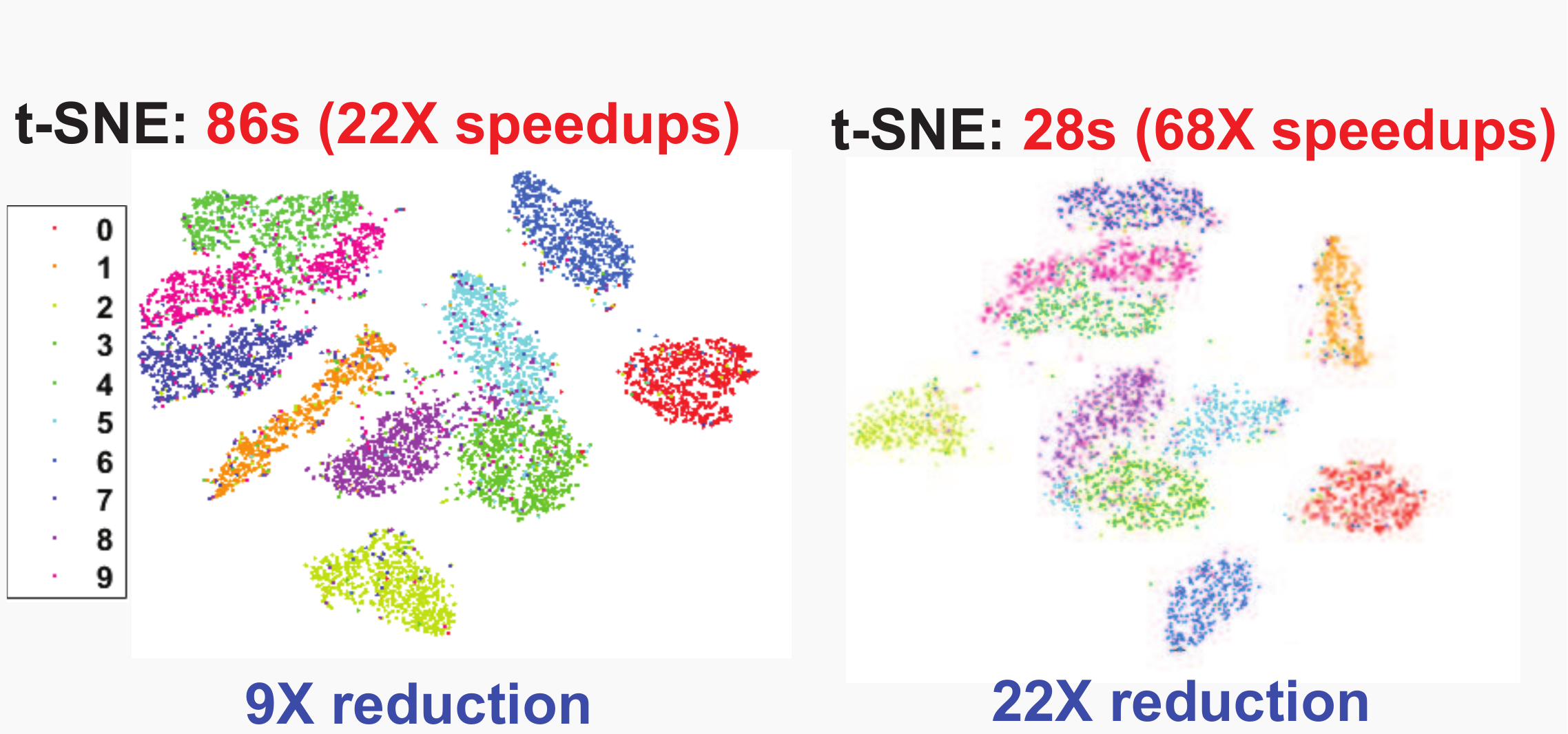}
	\caption{t-SNE visualization with original MNIST data set and data sets under different reduction ratios.\protect\label{fig:mnist2}}
\end{center}
\end{figure}

We also demonstrate the t-SNE visualization results obtained by leveraging spectrally-reduced NN graphs in Figures \ref{fig:usps} and \ref{fig:mnist2}. Our results show very clear cluster boundaries after spectral graph reduction, which retain the ones obtained from the original data sets, indicating very high-quality embedding results as well as significantly improved runtime performance.


\section{Conclusion}\label{sect:conclusion}
We propose  a scalable algorithmic framework for spectral reduction of large undirected graphs. The proposed method allows computing much smaller graphs while preserving the key spectrum of the original graph. To achieve truly scalable performance (nearly-linear complexity) for spectral graph reduction, we leverage recent similarity-aware spectral sparsification method, graph-theoretic algebraic multigrid (AMG) Laplacian solver and a novel constrained stochastic gradient descent (SGD) optimization approach in our spectral graph reduction algorithm. 

We show that the resulting spectrally-reduced graphs  can robustly preserve the first few nontrivial eigenvalues and eigenvectors of the original graph Laplacian.  In addition, the spectral graph reduction method has  been leveraged to develop much faster algorithms for multilevel spectral graph partitioning as well as t-distributed Stochastic Neighbor Embedding (t-SNE) of large data sets.  
We conducted extensive experiments using a variety of large graphs and data sets, and obtained very promising results. For instance, we are able to reduce the ``coPapersCiteseer" graph with $0.43$ million nodes and $16$ million edges to a much smaller graph with only $13K$ (32X fewer) nodes and $17K$ (950X fewer) edges in about 16 seconds; the spectrally-reduced graphs also allow  us to achieve up to  $1100X$ speedup for spectral graph partitioning and up to $60X$ speedup for t-SNE visualization of large data sets. 

\bibliographystyle{abbrv}
\bibliography{spectralredu,biggraph}

\begin{thebibliography}{10}

\bibitem{arthur2007k}
D.~Arthur and S.~Vassilvitskii.
\newblock k-means++: The advantages of careful seeding.
\newblock In {\em Proceedings of the eighteenth annual ACM-SIAM symposium on
  Discrete algorithms}, pages 1027--1035. Society for Industrial and Applied
  Mathematics, 2007.

\bibitem{bader2014benchmarking}
D.~A. Bader, H.~Meyerhenke, P.~Sanders, C.~Schulz, A.~Kappes, and D.~Wagner.
\newblock Benchmarking for graph clustering and partitioning.
\newblock In {\em Encyclopedia of Social Network Analysis and Mining}, pages
  73--82. Springer, 2014.

\bibitem{bader2012graph}
D.~A. Bader, H.~Meyerhenke, P.~Sanders, and D.~Wagner.
\newblock Graph partitioning and graph clustering.
\newblock In {\em 10th DIMACS Implementation Challenge Workshop}, 2012.

\bibitem{batson2012twice}
J.~Batson, D.~Spielman, and N.~Srivastava.
\newblock {Twice-Ramanujan Sparsifiers}.
\newblock {\em SIAM Journal on Computing}, 41(6):1704--1721, 2012.

\bibitem{benczur1996approximating}
A.~A. Bencz{\'u}r and D.~R. Karger.
\newblock Approximating st minimum cuts in {\~o} (n 2) time.
\newblock In {\em Proceedings of the twenty-eighth annual ACM symposium on
  Theory of computing}, pages 47--55. ACM, 1996.

\bibitem{briggs2000multigrid}
W.~L. Briggs, S.~F. McCormick, et~al.
\newblock {\em A multigrid tutorial}, volume~72.
\newblock Siam, 2000.

\bibitem{bulucc2016recent}
A.~Bulu{\c{c}}, H.~Meyerhenke, I.~Safro, P.~Sanders, and C.~Schulz.
\newblock Recent advances in graph partitioning.
\newblock In {\em Algorithm Engineering}, pages 117--158. Springer, 2016.

\bibitem{chen2011algebraic}
J.~Chen and I.~Safro.
\newblock Algebraic distance on graphs.
\newblock {\em SIAM Journal on Scientific Computing}, 33(6):3468--3490, 2011.

\bibitem{christiano2011flow}
P.~Christiano, J.~Kelner, A.~Madry, D.~Spielman, and S.~Teng.
\newblock Electrical flows, laplacian systems, and faster approximation of
  maximum flow in undirected graphs.
\newblock In {\em Proc. ACM STOC}, pages 273--282, 2011.

\bibitem{cohen2017almost}
M.~B. Cohen, J.~Kelner, J.~Peebles, R.~Peng, A.~B. Rao, A.~Sidford, and
  A.~Vladu.
\newblock Almost-linear-time algorithms for markov chains and new spectral
  primitives for directed graphs.
\newblock In {\em Proceedings of the 49th Annual ACM SIGACT Symposium on Theory
  of Computing}, pages 410--419. ACM, 2017.

\bibitem{davis2011matrix}
T.~Davis and Y.~Hu.
\newblock The university of florida sparse matrix collection.
\newblock {\em ACM Trans. on Math. Soft. (TOMS)}, 38(1):1, 2011.

\bibitem{defferrard2016convolutional}
M.~Defferrard, X.~Bresson, and P.~Vandergheynst.
\newblock Convolutional neural networks on graphs with fast localized spectral
  filtering.
\newblock In {\em Advances in Neural Information Processing Systems}, pages
  3844--3852, 2016.

\bibitem{donath1972algorithms}
W.~Donath and A.~Hoffman.
\newblock Algorithms for partitioning of graphs and computer logic based on
  eigenvectors of connections matrices.
\newblock {\em IBM Technical Disclosure Bulletin}, 15, 1972.

\bibitem{donath2003lower}
W.~E. Donath and A.~J. Hoffman.
\newblock Lower bounds for the partitioning of graphs.
\newblock In {\em Selected Papers Of Alan J Hoffman: With Commentary}, pages
  437--442. World Scientific, 2003.

\bibitem{zhuo:dac16}
Z.~Feng.
\newblock Spectral graph sparsification in nearly-linear time leveraging
  efficient spectral perturbation analysis.
\newblock In {\em Design Automation Conference (DAC), 2016 53nd ACM/EDAC/IEEE},
  pages 1--6. IEEE, 2016.

\bibitem{zhuo:dac18}
Z.~Feng.
\newblock Similarity-aware spectral sparsification by edge filtering.
\newblock In {\em Design Automation Conference (DAC), 2018 55nd ACM/EDAC/IEEE}.
  IEEE, 2018.

\bibitem{fiedler1973algebraic}
M.~Fiedler.
\newblock Algebraic connectivity of graphs.
\newblock {\em Czechoslovak mathematical journal}, 23(2):298--305, 1973.

\bibitem{fiedler1975property}
M.~Fiedler.
\newblock A property of eigenvectors of nonnegative symmetric matrices and its
  application to graph theory.
\newblock {\em Czechoslovak Mathematical Journal}, 25(4):619--633, 1975.

\bibitem{Fung:2011stoc}
W.~Fung, R.~Hariharan, N.~Harvey, and D.~Panigrahi.
\newblock A general framework for graph sparsification.
\newblock In {\em Proc. ACM STOC}, pages 71--80, 2011.

\bibitem{horn1990matrix}
R.~A. Horn, R.~A. Horn, and C.~R. Johnson.
\newblock {\em Matrix analysis}.
\newblock Cambridge university press, 1990.

\bibitem{karypis1995metis}
G.~Karypis and V.~Kumar.
\newblock Metis--unstructured graph partitioning and sparse matrix ordering
  system, version 2.0.
\newblock 1995.

\bibitem{karypis1998fast}
G.~Karypis and V.~Kumar.
\newblock A fast and high quality multilevel scheme for partitioning irregular
  graphs.
\newblock {\em SIAM Journal on scientific Computing}, 20(1):359--392, 1998.

\bibitem{kernighan1970efficient}
B.~W. Kernighan and S.~Lin.
\newblock An efficient heuristic procedure for partitioning graphs.
\newblock {\em The Bell system technical journal}, 49(2):291--307, 1970.

\bibitem{kolev2015note}
P.~Kolev and K.~Mehlhorn.
\newblock A note on spectral clustering.
\newblock {\em arXiv preprint arXiv:1509.09188}, 2015.

\bibitem{Kolla:2010stoc}
A.~Kolla, Y.~Makarychev, A.~Saberi, and S.~Teng.
\newblock Subgraph sparsification and nearly optimal ultrasparsifiers.
\newblock In {\em Proc. ACM STOC}, pages 57--66, 2010.

\bibitem{koren2003spectral}
Y.~Koren.
\newblock On spectral graph drawing.
\newblock In {\em International Computing and Combinatorics Conference}, pages
  496--508. Springer, 2003.

\bibitem{miller:2010focs}
I.~Koutis, G.~Miller, and R.~Peng.
\newblock {Approaching Optimality for Solving SDD Linear Systems}.
\newblock In {\em Proc. IEEE FOCS}, pages 235--244, 2010.

\bibitem{lee2014multiway}
J.~R. Lee, S.~O. Gharan, and L.~Trevisan.
\newblock {Multiway spectral partitioning and higher-order cheeger
  inequalities}.
\newblock {\em Journal of the ACM (JACM)}, 61(6):37, 2014.

\bibitem{Lee:2017}
Y.~T. Lee and H.~Sun.
\newblock {An SDP-based Algorithm for Linear-sized Spectral Sparsification}.
\newblock In {\em Proceedings of the 49th Annual ACM SIGACT Symposium on Theory
  of Computing}, STOC 2017, pages 678--687, New York, NY, USA, 2017. ACM.

\bibitem{lehoucq1997arpack}
R.~Lehoucq, D.~Sorensen, and C.~Yang.
\newblock Arpack users' guide: Solution of large scale eigenvalue problems with
  implicitly restarted arnoldi methods.
\newblock {\em Software Environ. Tools}, 6, 1997.

\bibitem{linderman2017clustering}
G.~C. Linderman and S.~Steinerberger.
\newblock Clustering with t-sne, provably.
\newblock {\em arXiv preprint arXiv:1706.02582}, 2017.

\bibitem{livne2012lean}
O.~Livne and A.~Brandt.
\newblock Lean algebraic multigrid ({LAMG}): Fast graph {L}aplacian linear
  solver.
\newblock {\em SIAM Journal on Scientific Computing}, 34(4):B499--B522, 2012.

\bibitem{pmlr-v80-loukas18a}
A.~Loukas and P.~Vandergheynst.
\newblock Spectrally approximating large graphs with smaller graphs.
\newblock In {\em Proceedings of the 35th International Conference on Machine
  Learning (ICML)}, pages 3237--3246, 2018.

\bibitem{maaten2008visualizing}
L.~v.~d. Maaten and G.~Hinton.
\newblock Visualizing data using t-sne.
\newblock {\em Journal of machine learning research}, 9(Nov):2579--2605, 2008.

\bibitem{naumov2016parallel}
M.~Naumov and T.~Moon.
\newblock Parallel spectral graph partitioning.
\newblock Technical report, NVIDIA Technical Report, NVR-2016-001, 2016.

\bibitem{newman2013community}
M.~E. Newman.
\newblock Community detection and graph partitioning.
\newblock {\em EPL (Europhysics Letters)}, 103(2):28003, 2013.

\bibitem{pavlopoulos2011}
G.~A. Pavlopoulos, M.~Secrier, C.~N. Moschopoulos, T.~G. Soldatos, S.~Kossida,
  J.~Aerts, R.~Schneider, and P.~G. Bagos.
\newblock Using graph theory to analyze biological networks.
\newblock {\em BioData mining}, 4(1):10, 2011.

\bibitem{peng2015partitioning}
R.~Peng, H.~Sun, and L.~Zanetti.
\newblock Partitioning well-clustered graphs: Spectral clustering works.
\newblock In {\em Proceedings of The 28th Conference on Learning Theory
  (COLT)}, pages 1423--1455, 2015.

\bibitem{saad2003iterative}
Y.~Saad.
\newblock {\em Iterative methods for sparse linear systems}, volume~82.
\newblock siam, 2003.

\bibitem{shi2000normalized}
J.~Shi and J.~Malik.
\newblock Normalized cuts and image segmentation.
\newblock {\em IEEE Transactions on pattern analysis and machine intelligence},
  22(8):888--905, 2000.

\bibitem{shuman2013emerging}
D.~I. Shuman, S.~K. Narang, P.~Frossard, A.~Ortega, and P.~Vandergheynst.
\newblock The emerging field of signal processing on graphs: Extending
  high-dimensional data analysis to networks and other irregular domains.
\newblock {\em IEEE Signal Processing Magazine}, 30(3):83--98, 2013.

\bibitem{Spielman:usg}
D.~Spielman and S.~Teng.
\newblock Nearly-linear time algorithms for graph partitioning, graph
  sparsification, and solving linear systems.
\newblock In {\em Proc. ACM STOC}, pages 81--90, 2004.

\bibitem{spielman2014sdd}
D.~Spielman and S.~Teng.
\newblock Nearly linear time algorithms for preconditioning and solving
  symmetric, diagonally dominant linear systems.
\newblock {\em SIAM Journal on Matrix Analysis and Applications},
  35(3):835--885, 2014.

\bibitem{spielman2010algorithms}
D.~A. Spielman.
\newblock Algorithms, graph theory, and linear equations in laplacian matrices.
\newblock In {\em Proceedings of the International Congress of Mathematicians},
  volume~4, pages 2698--2722, 2010.

\bibitem{spielman2011graph}
D.~A. Spielman and N.~Srivastava.
\newblock Graph sparsification by effective resistances.
\newblock {\em SIAM Journal on Computing}, 40(6):1913--1926, 2011.

\bibitem{spielman2011spectral}
D.~A. Spielman and S.-H. Teng.
\newblock Spectral sparsification of graphs.
\newblock {\em SIAM Journal on Computing}, 40(4):981--1025, 2011.

\bibitem{spielman2009note}
D.~A. Spielman and J.~Woo.
\newblock A note on preconditioning by low-stretch spanning trees.
\newblock {\em arXiv preprint arXiv:0903.2816}, 2009.

\bibitem{sutskever2013importance}
I.~Sutskever, J.~Martens, G.~Dahl, and G.~Hinton.
\newblock On the importance of initialization and momentum in deep learning.
\newblock In {\em International conference on machine learning}, pages
  1139--1147, 2013.

\bibitem{teng2016scalable}
S.-H. Teng.
\newblock Scalable algorithms for data and network analysis.
\newblock {\em Foundations and Trends{\textregistered} in Theoretical Computer
  Science}, 12(1--2):1--274, 2016.

\bibitem{van2014accelerating}
L.~Van Der~Maaten.
\newblock Accelerating t-sne using tree-based algorithms.
\newblock {\em The Journal of Machine Learning Research}, 15(1):3221--3245,
  2014.

\bibitem{von2007tutorial}
U.~Von~Luxburg.
\newblock A tutorial on spectral clustering.
\newblock {\em Statistics and computing}, 17(4):395--416, 2007.

\bibitem{wei1991ratio}
Y.-C. Wei and C.-K. Cheng.
\newblock Ratio cut partitioning for hierarchical designs.
\newblock {\em IEEE Transactions on computer-aided design}, 10(7):911--921,
  1991.

\bibitem{xue2009numerical}
F.~Xue.
\newblock {\em Numerical solution of eigenvalue problems with spectral
  transformations}.
\newblock PhD thesis, 2009.

\bibitem{zhiqiang:dac17}
Z.~Zhao and Z.~Feng.
\newblock A spectral graph sparsification approach to scalable vectorless power
  grid integrity verification.
\newblock In {\em Proceedings of the 54th Annual Design Automation Conference
  2017}, page~68. ACM, 2017.

\bibitem{zhiqiang:iccad17}
Z.~Zhao, Y.~Wang, and Z.~Feng.
\newblock {SAMG}: Sparsified graph theoretic algebraic multigrid for solving
  large symmetric diagonally dominant ({SDD}) matrices.
\newblock In {\em Proceedings of ACM/IEEE International Conference on
  Computer-Aided Design}, pages 601--606, 2017.

\end{thebibliography}

\appendix
\section{Spectral Graph Partitioning}\label{sec:graph_part}
Graph partitioning is one of the fundamental algorithmic operations, which can be applied to many fields  \cite{bulucc2016recent}, such as parallel processing, community detection in social networks \cite{newman2013community},  biological networks analysis \cite{pavlopoulos2011},  VLSI computer-aided design \cite{karypis1995metis}, etc. The goal of graph partitioning is to partition the vertices or edges of a graph into a number of disjoint sets without introducing too many connections between the sets. A variety of graph partitioning algorithms  has been proposed, from local heuristics like Kernighan-Lin \cite{kernighan1970efficient} to global methods such as spectral partitioning \cite{fiedler1975property, bulucc2016recent} and multilevel partitioning \cite{karypis1998fast}. Spectral partitioning, which was first noted in \cite{donath1972algorithms, donath2003lower, fiedler1973algebraic, fiedler1975property}, has become one of the most important methods for graph partitioning. 

Consider a weighted graph $G=(V,E_G,w_G)$ with vertex (node) set $V = \{v_1, \cdots, v_n\}$ denoting $n$ vertices in the graph, edge set $E_G$ representing weighted edges in the graph and $w_G$ denoting a weight function that assigns positive weight to all edges, that is $w_G(p,q)>0$ if there is an edge connecting node $v_p$ and node $v_q$, which can also be represented by $(p, q) \in E_G$.
Given a subset of vertices $S \subset V$ and its complement set $\bar{S} = V \backslash S$, the boundary between set $S$ and set $\bar{S}$ is defined as a set of edges $B(S, \bar{S}) \subset E_G$ such that one node of each edge is in set $S$ and the other node is in set $\bar{S}$: 
\begin{equation}\label{formula_boundary}
B(S) = \{(p,q): p \in S \wedge q \in \bar{S}\}.
\end{equation}
The cut between  $S$ and $\bar{S}$ can be defined as follows:
\begin{equation}\label{formula_cutvalue}
C(S, \bar{S}) = \sum_{(p,q)\in B(S)}w_G(p,q) = \mathrm{vol}(B(S)).
\end{equation}
The simplest idea of graph partitioning is to find a partition of the graph so that different partition sets are weakly connected (meaning the edges between different sets have low weights) while the interior of each set is strongly connected. The aim of graph partitioning is to find the set $S$ such that $C(S, \bar{S})$ can be minimized. However, in practice the solution of this min-cut problem is usually unacceptable due to the highly unbalanced partitioning results. For example, the resulted set $S$ may only have one individual vertex while $\bar{S}$ includes rest of the graph. Therefore, we also want the partitions to be reasonably balanced. To realize the minimum balanced cut of graph partitioning, two objective functions have been introduced: ratio cut $\rho(S)$ \cite{wei1991ratio} and normalized cut $\theta(S)$ \cite{shi2000normalized}, which have been defined as follows:
\begin{equation}\label{formula_cutratio}
\rho(S) = \underset{S}{\mathrm{min}} \frac{|V|C(S, \bar{S})}{|S||\bar{S}|} = \underset{S}{\mathrm{min}}\left(\frac{C(S, \bar{S})}{|S|} + \frac{C(S, \bar{S})}{|\bar{S}|}\right)
\end{equation}

\begin{equation}\label{formula_nratio}
\theta(S) = \underset{S}{\mathrm{min}} \frac{\mathrm{vol}(V)C(S, \bar{S})}{\mathrm{vol}(S)\mathrm{vol}(\bar{S})} =  \underset{S}{\mathrm{min}}\left( \frac{C(S, \bar{S})}{\mathrm{vol}(S)}+ \frac{C(S, \bar{S})}{\mathrm{vol}(\bar{S})}\right)
\end{equation}

where 
\begin{equation}\label{formula_S}
|S| := \text{the number of vertices in set S}
\end{equation}

\begin{equation}\label{formula_volS}
\mathrm{vol}(S) = \sum_{p \in S \wedge (p,q) \in E_G} w_G(p,q).
\end{equation}
Note that number of vertices (sum of edge weights) is used to measure the size of set $S$ for ratio cut $\rho(S)$ (normalized cut $\theta(S)$). In other words, the ratio cut $\rho(S)$ metric aims to balance the number of vertices for each set, while the normalized cut $\theta(S)$ metric aims to balance number of edges in each set. The ratio cut in (\ref{formula_cutratio}) and normalized cut in (\ref{formula_nratio}) can be generalized as follows for $k$-way partitioning problems \cite{naumov2016parallel, von2007tutorial}:
\begin{equation}\label{grc}
\rho(S_1, \cdots, S_k) =  \underset{S_1, \cdots, S_k}{\mathrm{min}}\sum_{i=1}^k{\frac{C(S_i, \bar{S_i})}{|S_i|}}
\end{equation}

\begin{equation}\label{gnc}
\theta(S_1, \cdots, S_k) =  \underset{S_1, \cdots, S_k}{\mathrm{min}}\sum_{i=1}^k{\frac{C(S_i, \bar{S_i})}{\mathrm{vol}(S_i)}},
\end{equation}
while the edge cut of $k$ partitions becomes
\begin{equation}\label{edgecut}
C(S_1, \cdots, S_k) =  \sum_{i=1}^k{C(S_i, \bar{S_i})}.
\end{equation}
Since the optimization problems of (\ref{grc}) and (\ref{gnc}) are NP-hard, spectral partitioning methods have been proposed for solving the relaxed optimization problems. It can be shown that the solution of the relaxed optimization problem (\ref{grc}) is the matrix of $U$ with first $k$ eigenvectors of the graph Laplacian as its columns vectors, whereas the solution to the relaxed optimization problem (\ref{gnc}) is the matrix of $U$ with the first $k$  eigenvectors of the normalized graph Laplacian \cite{shi2000normalized}. Detailed proof can be found in Section \ref{sect:k2} and Section \ref{sect:kk}.

Since we want to partition $V$ into $k$ sets based on the indicator matrix $U \in {\rm I\!R^{nxk}}$,  one straightforward way is to treat each row of the matrix $U$ as a point in a $k$ dimensional space and use clustering algorithms, like k-means \cite{arthur2007k} to identify the $k$ partitions. 

\subsection{Ratio cut and normalized cut for 2-way partitioning}\label{sect:k2}

Given the graph $G(V, E_G, w_G)$, the graph Laplacian $L_G$ is defined as follows:
\begin{equation}\label{formula_laplacian}
\mathbf{L}_G(i,j)=\begin{cases}
-w_G(i,j) & \text{ if } (i,j)\in E_G \\
\sum\limits_{(i,t)\in E_G}w_G(i,t) & \text{ if } (i=j) \\
0 & \text{ if } otherwise.
\end{cases}
\end{equation}
$L_G$ can also be represented as
\begin{equation}\label{formula_volS}
\mathbf{L_G = D_G - A_G},
\end{equation}
where $\mathbf{A}_G$ is the adjacency matrix of the graph and $\mathbf{D}_G$ is the diagonal matrix with each $i$-th diagonal element being the sum of all elements in that row of $\mathbf{A}_G$. To relate the ratio cut objective function with the unnormalized graph Laplacian, we first define the vector $\mathbf{z} = (z_1, z_2, \cdots, z_n)^T \in  {\rm I\!R^n}$  with entries noted as follows \cite{von2007tutorial, naumov2016parallel}:
\begin{equation}\label{formula_z}
z_i=\begin{cases}
\sqrt{(|\bar{S}|/|S|)} & \text{ if } v_i\in S \\
-\sqrt{(|S|/|\bar{S}|)} & \text{ if } v_i\in \bar{S} \\
\end{cases}
\end{equation}
Then we have 
\begin{equation*}\label{formula_zlz}
\begin{aligned}
\mathbf{z^TL_G z}  &=\frac{1}{2}\sum_{p,q=1}^n{w_G(p,q)(z_p-z_q)^2} \\
 & = C(S, \bar{S})\left(\frac{|S|+|\bar{S}|}{|S|}+\frac{|S|+|\bar{S}|}{|\bar{S}|}\right)\\
 & = |V|\cdot\frac{|V|C(S, \bar{S})}{|S||\bar{S}|}  = |V|\rho(S)
\end{aligned}
\end{equation*}
Given the all-one vector $\mathbf{1}$, the following  can be observed:
\begin{equation}\label{formula_sumz}
\mathbf{z}^T \mathbf{1} = \sum_{i=1}^n{z_i} = \sum_{i \in S}\sqrt{\frac{|\bar{S}|}{|S|}}-\sum_{i \in \bar{S}}\sqrt{\frac{|S|}{|\bar{S}|}} = 0,
\end{equation}

\begin{equation}\label{formula_zz}
\mathbf{z^T z} = \sum_{i=1}^n{z_i^2} = \sum_{i \in S}{\frac{|\bar{S}|}{|S|}}+\sum_{i \in \bar{S}}{\frac{|S|}{|\bar{S}|}} = n = |V|,
\end{equation}
which will lead to:
\begin{equation}\label{formula_rho}
\rho(S) =  \frac{\mathbf{z^TL_G z}}{\mathbf{z^T z}}.
\end{equation}


Since the values of $z_i$ are restricted to the two particular values, this optimization problem is NP hard. Spectral partitioning   relaxes the constrains and allows $\mathbf{z}$ to take any real values. 

According to the \textbf{Courant-Fischer theorem } \cite{horn1990matrix}, the solution to the relaxed optimization problem is the eigenvector of $\mathbf{L_G}$ associated with the smallest non-zero eigenvalue. 
Once the solution vector $\mathbf{z}$ is computed, a partition can be obtained by converting the real-valued vector $\mathbf{z}$ to a 
discrete vector containing only $0$ and $1$ as the indicators for partitioning purpose. For example, one simple way is to use the sign of the vector $\mathbf{z}$ to partition the graph so that $v_i \in S$ if $z_i>0$, otherwise $v_i \in \bar{S}$. Similar analysis can be performed for normalized cut by setting the vector $\mathbf{z}$ to be: 
\begin{equation}\label{formula_nz}
z_i=\begin{cases}
\sqrt{(\mathrm{vol}(\bar{S})/\mathrm{vol}(S))} & \text{ if } v_i\in S \\
-\sqrt{(\mathrm{vol}(S)/\mathrm{vol}(\bar{S}))} & \text{ if } v_i\in \bar{S} \\
\end{cases}
\end{equation}
which leads to:
\begin{equation}\label{formula_nz1}
\mathbf{z^TL_Gz} = \mathrm{vol}(V)\theta(S)
\end{equation}

\begin{equation}\label{formula_nz2}
\mathbf{(D_Gz)^T}\mathrm{1} = 0
\end{equation}

\begin{equation}\label{formula_nz3}
\mathbf{z^TD_Gz} = \mathrm{vol}(V)
\end{equation}


By relaxing the vector $\mathbf{z}$ to take arbitrary real values, we can show that the solution to this relaxed optimization problem is the eigenvector associated to the second smallest eigenvalue to the generalized eigenvalue problem of 
\begin{equation}\label{formula_nz4}
\mathbf{L_Gu} = \lambda \mathbf{D_Gu}.
\end{equation}

\subsection{Ratio cut and normalized cut for k-way partitioning} \label{sect:kk}
It is similar to previous analysis when relaxing the ratio cut and normalized cut minimizations to the general $k$ partitions. Given a partition of $V$ into $k$ sets, we define $k$ indicator vectors $\mathbf{m_j} = (m_{1,j}, \cdots, m_{n, j})^T$ with $j=1, \cdots, k$ such that

\begin{equation}\label{formula_m}
m_{i,j}=\begin{cases}
1/\sqrt{|S_j|} & \text{ if } v_i\in S_j \\
0 & \text{ otherwise  } \\
\end{cases}
\end{equation}

where $i=1, \cdots, n;\;\; j=1, \cdots, k$.

Indicator matrix $\mathbf{U}$ can be defined with the $k$ vectors so that $\mathbf{U} = [\mathbf{m_1}, \cdots, \mathbf{m_k}]$. Note that columns in $\mathbf{U}$ are orthogonal to each other, that is $\mathbf{U^TU =I}$.   We   also note that

\begin{equation}\label{gnc1}
\mathbf{m_j^TL_Gm_j} =  (\mathbf{U^TL_GU})_{jj} = \frac{C(S_j, \bar{S_j})}{|S_j|}
\end{equation}

By substituting (\ref{gnc1}) to (\ref{grc}) we can get

\begin{equation}\label{gnc11}
\rho(S_1, \cdots, S_k) = \sum_{j=1}^k{\mathbf{m_j^TL_Gm_j}} = \sum_{j=1}^k{(\mathbf{U^TL_GU})_{jj}} =Tr(\mathbf{U^TL_GU})
\end{equation}

where $Tr(\cdot)$ is the trace of a matrix. By relaxing the entries of indicator matrix $\mathbf{U}$ to be arbitrary real values, the optimization problem in (\ref{grc}) becomes

\begin{equation*}\label{formula_kgnc}
\begin{aligned}
&\rho(S) = \underset{\mathbf{U} \in {\rm I\!R^{nxk}}}{\mathrm{min}} Tr(\mathbf{U^TL_GU})  \text{  \;\; subject to: } \mathbf{U^TU = I}.
\end{aligned}
\end{equation*}
According to the \textbf{Courant-Fischer} theorem, the solution to this optimization problem is the matrix of $\mathbf{U}$ with first $k$ eigenvectors of $\mathbf{L_G}$ as its columns.

Similarly, we can choose the entries of indicator matrix $\mathbf{U}$ as follows:

\begin{equation}\label{formula_m1}
m_{i,j}=\begin{cases}
1/\sqrt{\mathrm{vol}(S_j)} & \text{ if } v_i\in S_j \\
0 & \text{ otherwise  }.
\end{cases}
\end{equation}
We  observe that $\mathbf{U^TD_GU = I}$, and $\mathbf{m_j^TL_Gm_j} = C(S_j, \bar{S_j})/\mathrm{vol}(S_j)$. By relaxing $\mathbf{U}$ to take arbitrary real values, we can reformulate the optimization problem in $(\ref{gnc})$ as 

\begin{equation*}\label{formula_kgnc1}
\begin{aligned}
&\theta(S) = \underset{\mathbf{U} \in {\rm I\!R^{nxk}}}{\mathrm{min}} Tr(\mathbf{U^TL_GU})  \text{  \;\; subject to: } \mathbf{U^TD_GU = I}.
\end{aligned}
\end{equation*}
According to the \textbf{Courant-Fischer} theorem, the solution to this optimization problem is the matrix of $\mathbf{U}$ with first $k$ generalized eigenvectors of $\mathbf{L_Gu}=\lambda \mathbf{D_Gu}$ as its columns \cite{shi2000normalized}.
\section{t-Distributed Stochastic Neighbor Embedding}
t-Distributed Stochastic Neighbor Embedding (t-SNE) \cite{maaten2008visualizing, van2014accelerating} is a nonlinear dimensionality reduction method which is well suited for visualization of data. The goal of t-SNE is to learn a mapping from the high-dimensional space to a low-dimensional space with desired number of dimensionality in such a way that similar data points are mapped to nearby locations and dissimilar data points are mapped to distant locations. To this end, t-SNE converts the euclidean distances between data points in high-dimensional space into conditional probability as follows:

\begin{equation}\label{formula_cond prob}
{P_{j\mid i}} =  {{{\frac{exp(-\frac{{\|x_i-x_j\|}^2}{2{\sigma_i}^2})}{\sum_{k \neq i}exp(-\frac{{\|x_i-x_k\|}^2}{2{\sigma_i}^2})}}}},~~{P_{i\mid i}} = 0,
\end{equation}
where $\sigma_i$ denotes the variance of the Gaussian distribution that is centered at $x_i$. The joint probability is defined by symmetrizing a pair of conditional probabilities as follows:
\begin{equation}\label{formula_joint prob high}
{P_{ij}} =  {{{\frac{P_{j\mid i}+P_{i\mid j}}{2{N}}}}},
\end{equation}
 t-SNE uses this joint probability to measure the similarity between two data points $x_i$ and $x_j$ in high-dimensional space. Denoting the corresponding points in low-dimensional space  by $y_i$ and $y_j$, respectively, then the similarity between them is measured by the following joint probability using the Cauchy kernel:
\begin{equation}\label{formula_joint prob low}
{q_{ij}} =  {{{\frac{(1+{{\|y_i-y_j\|}^2})^{-1}}{\sum_{k \neq l}(1+{{\|y_k-y_l\|}^2})^{-1}}}}},~~~{q_{ii}} = 0.
\end{equation}
t-SNE uses Kullback-Leibler (KL) divergence to measure the faithfulness of the embedding. The cost function is defined as the sum of KL divergence over all pairs of data points in the data set:
\begin{equation}\label{formula_KL}
{C} = KL({P\parallel Q})= \sum\limits_{i \neq j} {{p_{ij}}} \log{\frac{p_{ij}}{q_{ij}}}.
\end{equation}
The embedding points in low-dimensional space \{$y_1,...,y_N$ \} are determined by minimizing the KL cost function. Typically, starting with a random initialization, the cost function is minimized using gradient descent method with the following gradient:
\begin{equation}\label{formula_gd}
\frac{\partial C}{\partial y_i}= 4\sum\limits_{i \neq j}(p_{ij}-{q_{ij}}){q_{ij}}Z(y_i-y_j),
\end{equation}
where the constant normalization term Z is given by:
\begin{equation}\label{formula_normalization}
Z={\sum_{k \neq l}(1+{{\|y_k-y_l\|}^2})^{-1}}.
\end{equation}
It should be noted that as the size of data set grows, the convergence rate will usually slow down \cite{maaten2008visualizing, van2014accelerating}.
Computing gradients is very time-consuming, since it is an N-body problem that has a complexity of $O(N^2)$. By splitting the gradient into two parts, we have:
\begin{equation}\label{formula_gd_split}
\begin{array}{l}
\frac{\partial C}{\partial y_i}= 4\sum\limits_{i \neq j}p_{ij}{q_{ij}}Z(y_i-y_j)-4\sum\limits_{i \neq j}{q_{ij}^2}Z(y_i-y_j)\\=4F_{attr,i}-4F_{rep,i},
\end{array}
\end{equation}
where $F_{attr,i}$ denotes the sum of attractive force acting on data point $i$ and the $F_{rep,i}$ denotes the sum of repulsive force acting on data point $i$. Both forces are due to the rest of the data points. The position of each data point after embedding is determined by the net force acting on it. 

In recent years, due to the prevalence of high-dimensional data, the t-SNE algorithm has become the most effective visualization tool due to its capability of performing dimensionality reduction in such a way that the similar data points in high-dimensional space are embedded to nearby locations in low-dimensional space of two or three dimensions with high probability. However, t-SNE is limited in its applicability to large real-world data sets due to the high computational complexity. In practice, the standard t-SNE can not even been applicable to data sets with more than $10,000$ data points \cite{maaten2008visualizing}. Thus, there is a pressing need to develop acceleration techniques for t-SNE algorithm that can be adapted for visualizing large-scale data sets. In the past decade, substantial effort has been made to reduce the computational cost of t-SNE. For example,  tree-based algorithms have been proposed to accelerate the computation of the gradient in t-SNE \cite{van2014accelerating}, which however has no theoretical guarantee of the solution quality.
\end{document}